\documentclass{JFM-FLM_Au}

\usepackage{bm}
\usepackage{array}
\usepackage{booktabs}
\usepackage{multirow}

\usepackage{tikz}
\usetikzlibrary{patterns}
\usepackage{caption}
\usepackage{subcaption}

\usepackage{amsmath}

\lefttitle{K. Zhu, N. Bempedelis and K. Steiros}
\righttitle{Journal of Fluid Mechanics}

\title{Data-driven identification of multiscale self-similarity and asymptotic matching}

\author{Kaixin Zhu\aff{1}, Nikos Bempedelis\aff{2} \and Konstantinos Steiros\aff{1} }

\affiliation{\aff{1}Department of Aeronautics, Imperial College London, London SW7 2AZ, UK
\aff{2}School of Engineering and Materials Science, Queen Mary University of London, London E1 4NS, UK}

\corresau{Kaixin Zhu, k.zhu1@imperial.ac.uk; Konstantinos Steiros, k.steiros@imperial.ac.uk}

\begin{document}
\maketitle

\begin{abstract}
A central problem in fluid mechanics is the identification of self-similarity, which reveals the underlying scaling behaviour of a flow. Recently, a data-driven framework proposed by Bempedelis \textit{et al.} (2025 \textit{J. Fluid Mech.}, vol. 1020, A11) has enabled the extraction of single-scale self-similarity from data, without prior knowledge of the governing equations. However, many physical systems are inherently multiscale and therefore require more general approaches.
Building on this foundation, we develop an algorithm for the systematic identification of multiscale self-similarity. The algorithm is first applied to two canonical fluid-mechanical problems: turbulent channel flow and late-stage homogeneous decaying turbulence characterised by classical dissipation laws. In both cases, the algorithm successfully identifies inner and outer self-similarity from numerical data and recovers the corresponding similarity expressions.
In the intermediate region of both problems, the algorithm identifies similarity expressions that are independent of both the inner and outer scales, providing a novel data-driven route to recovering the corresponding scaling laws: the logarithmic law in turbulent channel flow and the $-5/3$ law in late-stage homogeneous decaying turbulence.
Moreover, the algorithm uncovers corrections to both laws: one linked to the power-law scaling proposed by Barenblatt (1993 \textit{J. Fluid Mech.}, vol. 248, 513--520), and the other yielding an improved approximation of the energy spectrum. The algorithm is then applied to early-stage homogeneous decaying turbulence, characterised by the nonclassical dissipation law. It identifies the same inner similarity as in the late stage, but different outer similarity expressions.

\end{abstract}

\begin{keywords}
Authors should not enter keywords on the manuscript, as these must be chosen by the author during the online submission process and will then be added during the typesetting process (see \href{https://www.cambridge.org/core/journals/journal-of-fluid-mechanics/information/list-of-keywords}{Keyword PDF} for the full list).  Other classifications will be added at the same time.
\end{keywords}




\section{Introduction}\label{sec-intro}

Fluid systems are inherently multiscale, spanning a wide range of interacting spatial and temporal scales. For example, in boundary-layer flow at high Reynolds numbers, a thin viscous layer (small wall-normal scale) develops along a long plate (large streamwise scale); in turbulent wakes, large-scale vortex shedding (large eddy scale) coexists with small-scale dissipative motions (small eddy scale); and in turbulent combustion, rapid chemical reactions (fast time scale) are coupled with slow transport processes (slow time scale). This multiscale nature continues to challenge both simulations and experiments: resolving the full spectrum of scales in numerical simulations demands immense computational resources \citep{moin1998direct}, whereas experimental measurements are typically limited to a subset of scales due to finite spatial and temporal resolution \citep{lavoie2007spatial}.

A common approach to simplifying such complex systems is to uncover their underlying self-similarity.
When the equations governing a system are known and closed, the identification of self-similarity can reduce the dimensionality of the problem, enabling more tractable numerical solutions or analytical and semi-analytical solutions. Even in the absence of governing equations, or when they are not closed, identifying self-similarity remains valuable, as it can reveal invariant scaling relationships that reflect the underlying physical mechanisms and simplify the description of the system.
Conversely, the breakdown of self-similarity may signal the emergence of additional physical processes or interactions not captured by the current framework, thereby revealing model limitations and motivating refinement. 
The identification of self-similarity is therefore a central step in the analysis and modelling of complex fluid systems.
Self-similarity has been successfully identified in a variety of fluid-mechanical contexts, for example, in solutions to Stokes problems \citep{stokes2009effect,pritchard2011stokes}, velocity profiles of laminar boundary-layer flows \citep{blasius1908,falkner1931solutions,schlichting2016,zhu2022effects}, and the mean velocity and Reynolds stresses of turbulent shear flows \citep{townsend1976structure,pope2000turbulent}. Self-similarity also underpins the closure of turbulence models. By treating small-scale eddies as statistically universal, turbulence can be modelled without explicitly resolving all scales. This assumption is widely employed in classical turbulence modelling frameworks, including Reynolds-averaged Navier--Stokes models and large-eddy simulation \citep{durbin2018some}.

The identification of self-similarity relies on a combination of mathematical analysis and physical intuition.
The theory of Lie groups \citep{cantwell2002introduction,oberlack2001unified} and dimensional analysis \citep{barenblatt1972selfsimilar,barenblatt2003scaling} can be applied when a mathematical formulation is available. In the absence of such a formulation, dimensional analysis can still be applied, but its success depends on the choice of the relevant variables and parameters. In some cases, self-similarity may also be inferred directly from data on the basis of physical intuition.
However, these approaches are largely restricted to problems with relatively simple mathematical descriptions and well-understood physical mechanisms, or they depend strongly on the practitioner's skills.
Recently, data-driven approaches have emerged as a powerful means of discovering symmetries \citep{barenboim2021symmetry,liu2022machine,otto2025unified} and self-similarity \citep{bhattacharjee2001measure,xie2022data,bakarji2022dimensionally,watanabe2025data,yuan2025dimensionless,bempedelis2025extracting}.
Despite this progress, these methods typically consider the full data field and seek a single self-similar scaling that captures the dominant behaviour across all scales. This is effective when a single self-similar regime governs the data, but in many realistic systems the data exhibit multiple self-similar regimes across scales, each associated with distinct scaling laws and reflecting different physical mechanisms. Moreover, rather than being cleanly separated so that existing methods could be applied to each regime individually, these regimes may overlap and coexist over intermediate ranges, making it difficult for existing methods to capture their scaling behaviour and mechanisms.
One example is wall-bounded turbulent shear flow, where different self-similar behaviours exist in the near-wall and outer regions, while their overlap gives rise to the logarithmic law \citep{tennekes1972first,pope2000turbulent}. 

\citet{bempedelis2025extracting} recently proposed a data-driven framework for extracting self-similarity without prior knowledge of the governing equations. 
The framework successfully identifies single-scale self-similarity from both numerical and experimental data, as demonstrated by its discovery of physical laws in several fluid-mechanics problems.
However, when applied to homogeneous decaying turbulence, the framework fails to reveal the inner--outer similarities, including the $-5/3$ scaling law in the overlap region, due to its inability to treat multiscale dynamics. Since multiscale similarity and the associated overlapping scaling laws are fundamental to turbulence, arising in turbulent wall-bounded flows \citep{pope2000turbulent}, thermally stratified boundary layers \citep{kader1990mean}, and vortex flows \citep{steiros2017transient,vanKuik2017joukowsky}, this limitation underscores the need for a more general framework.
In this paper, we extend the framework of \citet{bempedelis2025extracting} to enable the identification of multiscale self-similarity, thereby elucidating the underlying mechanisms of multiscale problems.

The remainder of the paper is organised as follows. In \S\ref{sec-methodology}, we describe the data-driven methodology for extracting multiscale self-similarity. In \S\ref{sec-results}, we apply the method to three problems in fluid mechanics: turbulent channel flow, late-stage homogeneous decaying turbulence, and early-stage homogeneous decaying turbulence. In \S\ref{sec-sum}, we summarise the main conclusions and suggest directions for future work.

\section{Methodology}\label{sec-methodology}
\subsection{Two-step method for the identification of self-similarity}

We consider the numerical computation or experimental measurement of a quantity of interest, $q(s,t)$, where $s$ and $t$ are governing parameters. The quantity is evaluated at discrete stations $t_j$ over the $s$ range from $s_j^0$ to $s_j^1$, with $j=1,\ldots,N_t$, where the superscripts 0 and 1 represent the lower and upper limits of the range, respectively, and $N_t$ denotes the total number of stations.
Following \citet{bempedelis2025extracting}, we propose a two-step method for identifying self-similarity over a selected range of $s$ at a reference station and corresponding matched ranges from other stations. 

\subsubsection{Step 1. Search for transformation functions}

\begin{figure}
  \centerline{\includegraphics[width=\textwidth]{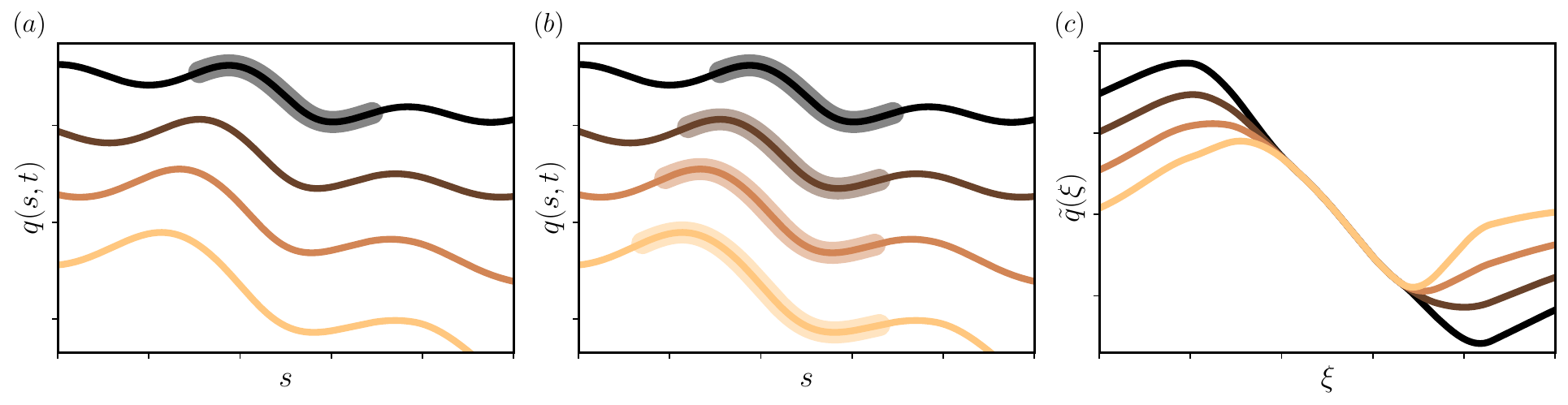}}
  \caption{$(a)$ A specific range of $s$ from a reference station $t_r$ (black curve) is selected. $(b)$ The corresponding matched ranges from other stations $t_j$ are identified by solving an optimisation problem. $(c)$ If self-similarity exists, the matched ranges collapse onto a single curve $\tilde{q}(\xi)$ under the transformation $q(s,t) \to \tilde{q}(\xi)$.}
  \label{fig-method}
\end{figure}

In Step 1, a specific range of $s$ from a reference station $t_r$ is selected, namely $[s_r^l,s_r^u]\in[s_r^0,s_r^1]$, and an optimisation problem is formulated to identify the corresponding matched range, $[s_j^l,s_j^u]$, from other stations ($j\neq r$) that aligns with the reference range, together with the corresponding transformation functions. The superscripts $l$ and $u$ denote the lower and upper bounds of the matched range, respectively. 
If self-similarity exists, a transformation $q(s,t) \to \tilde{q}(\xi)$, with $\xi = \xi(s,t)$, can be identified such that the matched ranges from all stations collapse onto a single curve $\tilde{q}(\xi)$. The similarity variables $\tilde{q}$ and $\xi$ are dimensionless. Figure \ref{fig-method} illustrates the above procedure.

We express the similarity variables as a combination of elementary dilation and translation transformations,
\begin{gather}
  \xi_j = \alpha(t_j)s + \beta(t_j),\label{eq-xi}\\
  \tilde{q}_j = \gamma(t_j) q(s,t_j) + \delta(t_j),\label{eq-qtilde}
\end{gather}
where $\alpha$ and $\gamma$ are dilation functions, and $\beta$ and $\delta$ are translation functions. For the reference station, we set $\alpha(t_r)=\gamma(t_r)=1$ and $\beta(t_r)=\delta(t_r) = 0$. Unlike approaches that express similarity variables as power-law monomials of physical variables and search for the corresponding exponents \citep{bhattacharjee2001measure,watanabe2025data}, this formulation admits more general similarity mappings. In particular, it can be readily extended to higher-order problems involving transformations such as rotations and shearing.

An optimisation problem is then formulated to identify the values of $\alpha, \beta, \gamma$ and $\delta$ for the stations $t_j$ ($j\neq r$) that minimise the distance in the transformed quantity $\tilde{q}$ across all stations:
\begin{gather}
  \operatorname*{arg\,min}_{\bm{w}} J(\bm{w}), \label{opt}
\end{gather}
where $\bm{w} = (\alpha, \beta, \gamma, \delta)$ and $J(\bm{w})$ is the objective function defined as
\begin{gather}
  J(\bm{w}) = \frac{1}{N_\xi}\frac{1}{N_t}\sum_{j>k}^{N_t} \sum_{k=1}^{N_t} \left\| \frac{\tilde{q}_j - \tilde{q}_k}{(\tilde{q}_j + \tilde{q}_k)/2} \right\|_2^2. \label{loss}
\end{gather}
The $l_2$-norms, $\|\cdot\|_2^2$, are computed over the overlap interval $[\xi^l,\xi^u]$ between different stations in the transformed coordinate $\xi$. The values of $\tilde{q}$ on this interval are interpolated from the input data, with the number of grid points $N_\xi$ chosen arbitrarily. The objective function \eqref{loss} is normalised by the total number of stations $N_t$ and grid points $N_\xi$, ensuring independence from both. For stations with $j\neq r$, the following constraints ensure that the transformed range of each station covers the transformed reference interval:
\begin{gather}
  \xi_j^l \le \xi_r^l, \quad  \xi_j^u \ge \xi_r^u.
\end{gather}
The matched range $[s_j^l,s_j^u]$ for station $t_j$ ($j\neq r$) is then determined a posteriori by applying the inverse transformation to the overlap interval $[\xi^l,\xi^u]$ using \eqref{eq-xi}.

\subsubsection{Step 2. Analytic form of the transformation}
With the discrete values of the transformation functions $\alpha, \beta, \gamma$ and $\delta$ known at each station $t_j$ from Step 1, symbolic regression is applied in Step 2 to identify their analytic forms.  Symbolic regression is a machine learning technique that searches for symbolic expressions that best fit the data, given a set of candidate constants, functions, operations, and state variables. A variety of methods have been developed for this purpose, including genetic programming \citep{dubcakova2011eureqa}, sparse regression \citep{brunton2016discovering}, and physics-inspired approaches \citep{udrescu2020aifeynman}.

This step of the algorithm can be implemented with any symbolic regression method. In this paper, we employ the custom regression algorithm developed by \citet{bempedelis2025extracting}. 
The identified analytic forms of $\alpha, \beta, \gamma$ and $\delta$ are then used to construct the similarity variables $\xi$ and $\tilde{q}$ through \eqref{eq-xi} and \eqref{eq-qtilde}.

\begin{figure}
  \centerline{\includegraphics[width=0.7\textwidth]{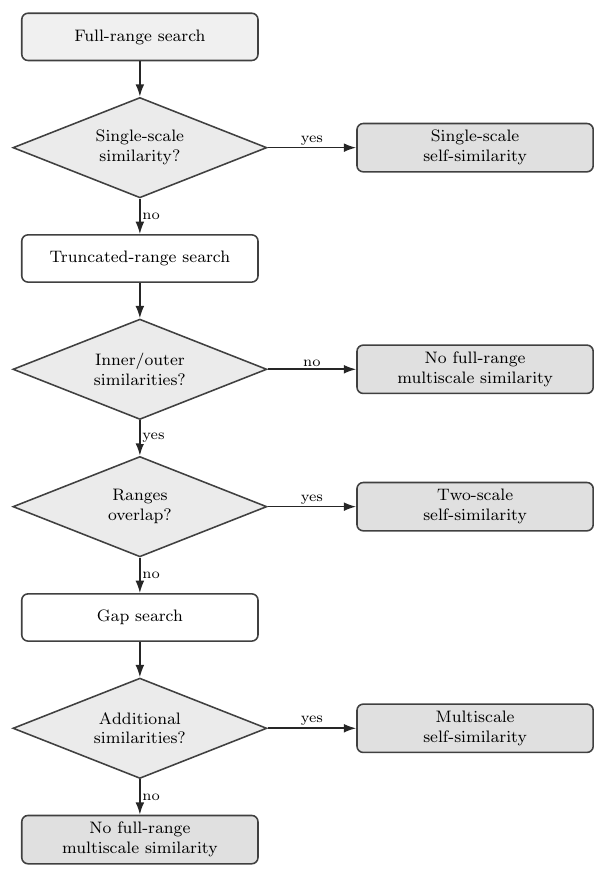}}
  \caption{Workflow for identifying multiscale self-similarity.}
  \label{fig-flowchart}
\end{figure}

\subsection{Workflow for identifying multiscale self-similarity}

The method described above enables the extraction of self-similarity in $q(s,t)$ over a specified range of $s$ for a reference station $t_r$. Based on this approach, we propose a workflow to systematically identify possible multiscale self-similarities, as illustrated in figure \ref{fig-flowchart}. The procedure is as follows.

\begin{figure}
  \centerline{\includegraphics[width=0.85\textwidth]{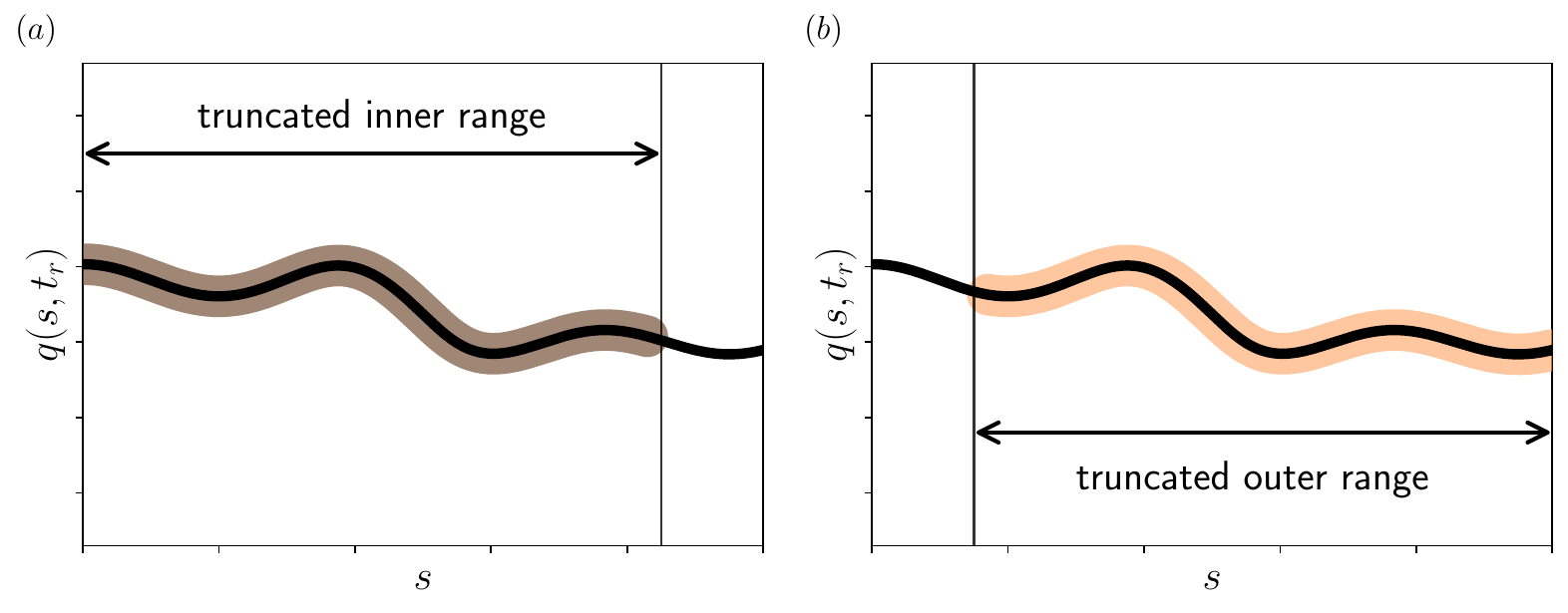}}
  \caption{Truncated $s$-range of the reference station $t_r$ for identifying inner and outer self-similarity. Black lines: the profile of the reference station. Shaded regions: the truncated inner and outer $s$-ranges.}
  \label{fig-truncated-range}
\end{figure}

\begin{enumerate}
  \item \textit{ Search for single-scale self-similarity.}  
  We first apply the algorithm over the full range of $s$ across all stations. If self-similarity is identified over this entire range, the system is considered to exhibit single-scale self-similarity.

  \item \textit{ Search for two-scale self-similarity.}  
  If single-scale self-similarity is not identified, we then examine the possibility of two-scale self-similarity, comprising inner and outer self-similarity. This is achieved by progressively truncating the $s$-range of the reference station from both ends and applying the algorithm to each truncated range, as illustrated in figure \ref{fig-truncated-range}.
  If both inner and outer self-similarity are identified, and the two ranges overlap, the system is considered to exhibit two-scale self-similarity.

  \item \textit{ Search for multiscale self-similarity.}  
  If the identified inner and outer self-similar ranges do not overlap, the same procedure is applied to the gap between them to search for additional self-similar ranges. This process is repeated until all potential self-similar ranges are identified or no additional self-similarity can be detected.
\end{enumerate}

The existence of self-similarity is determined by evaluating the objective function $J(\bm{w})$ defined in \eqref{loss}. Two criteria are used. The first is a fixed-threshold criterion, where self-similarity is identified if $J$ falls below a prescribed value, with the threshold chosen according to the noise level in the data. The threshold is set to 0.001 in this paper.
A similar criterion was adopted by \citet{liu2022machine}, who presented a machine learning method for discovering hidden symmetries following a similar two-step procedure. 
The second criterion is used specifically in the search for inner and outer self-similar ranges when the noise level is high, where self-similarity is identified when $J$ attains a global minimum as the range of $s$ is progressively truncated.

The proposed workflow is designed to identify systems exhibiting single-, two-, and multiscale self-similarity covering the full $s$ range. When self-similarity is restricted to a local region, for example, around the middle of the $s$ range, the flexibility of the two-step framework also allows the corresponding self-similar range to be extracted. In such a case, the algorithm is applied to the selected local range of the reference station.

In the following section, we apply this workflow to three problems in fluid mechanics: turbulent channel flow, late-stage homogeneous decaying turbulence, and early-stage homogeneous decaying turbulence.

\section{Results}\label{sec-results}
\subsection{Turbulent channel flow}\label{sec-turbulent_channel}

Wall-bounded turbulence provides an example of two-scale self-similarity, where the inner region near the wall is characterised by the viscous length scale $\delta_\nu=\nu/u_\tau$, with $\nu$ being the kinematic viscosity and $u_\tau=\sqrt{\tau_w/\rho}$ the friction velocity defined in terms of the wall shear stress $\tau_w$ and fluid density $\rho$, and the outer region by a macroscopic length scale. Here, we focus on turbulent channel flow, where the outer length scale is typically taken as the channel half-height $\delta$. The flow is commonly considered to be completely described by $\rho$, $\nu$, $\delta$, and $u_\tau$ \citep{pope2000turbulent}. Based on dimensional analysis, the streamwise mean velocity gradient $\mathrm{d}\bar{U}/\mathrm{d}y$ can then be expressed as
\begin{gather}
  \frac{\mathrm{d}\bar{U}}{\mathrm{d}y} = \frac{u_\tau}{y} \Phi\left(\frac{y}{\delta_\nu}, \frac{y}{\delta}\right),
\end{gather}
where $\bar{U}$ is the mean streamwise velocity, $y$ represents the wall-normal distance from the channel wall, and $\Phi$ is a universal nondimensional function.

\subsubsection{Single-scale self-similarity} \label{sec-ch-single}

\begin{figure}
  \centerline{\includegraphics[width=\textwidth]{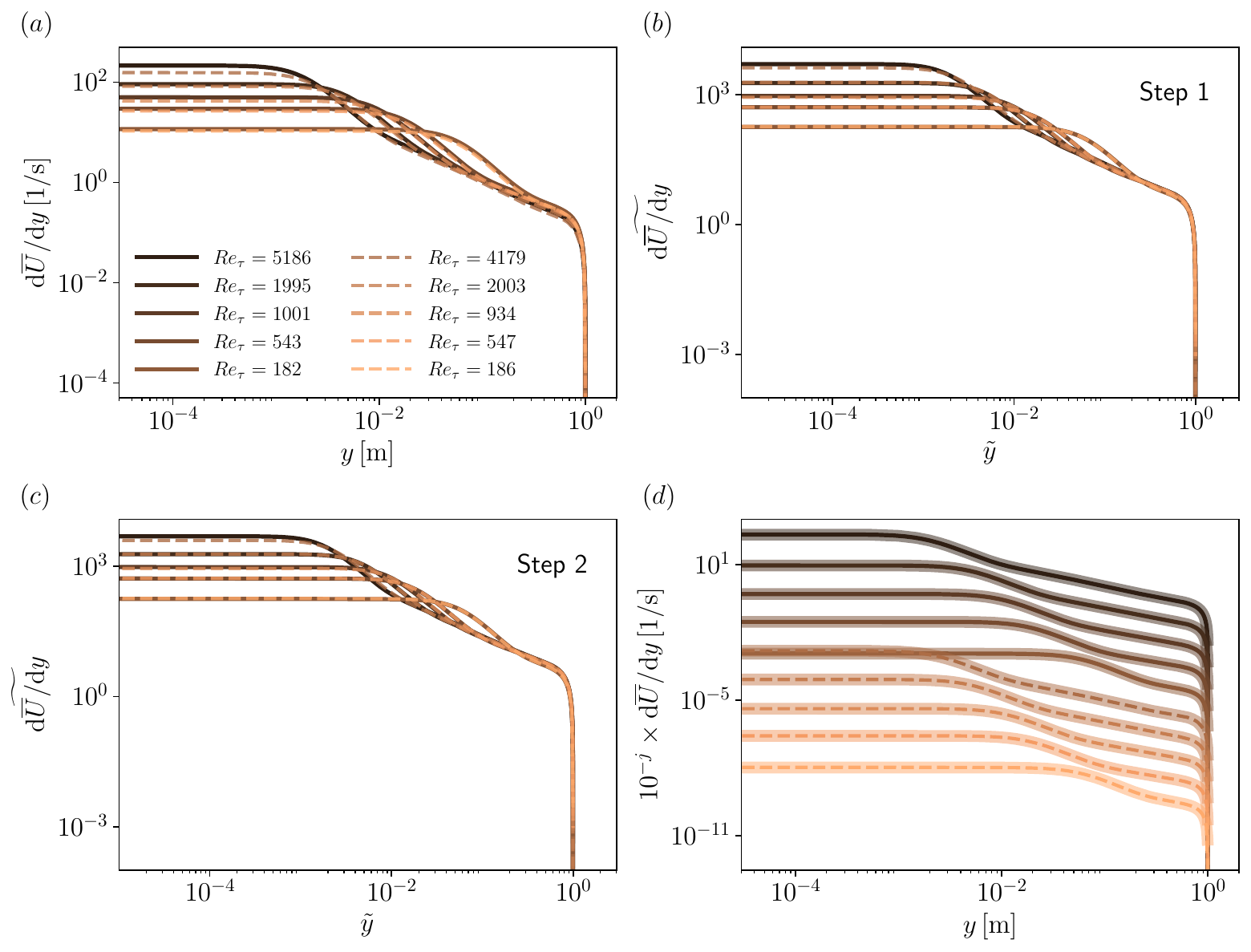}}
  \caption{
  $(a)$ Mean velocity gradient profiles at different Reynolds numbers $Re_\tau$, taken from two DNS datasets: \citet{lee2015direct} (solid lines) and Jim{\'e}nez and co-workers \citep{lozano-duran2014effect, hoyas2006scaling, delalamo2004scaling} (dashed lines). The same line styles are used in figures \ref{fig-ch_inner_1}--\ref{fig-ch_ab_y_4}, \ref{fig-ch_inner_2}--\ref{fig-ch_ab_y_6}.
  $(b)$ Rescaled mean velocity gradient profiles obtained using the expression \eqref{eq-ch-one-scale-step1}, identified in Step 1.
  $(c)$ Rescaled mean velocity gradient profiles obtained using the expression \eqref{eq-ch-one-scale-1}, identified in Step 2.
  $(d)$ The corresponding collapsed range (shaded) of each profile. The profiles are premultiplied by $10^{-j}$ for visual separation. Indices $j$ = 0 to 4 correspond to data from \citet{lee2015direct}, while indices $j$ = 5 to 9 correspond to those from Jim{\'e}nez and co-workers, each ordered by decreasing $Re_\tau$.
  }
  \label{fig-ch_one_scale}
\end{figure}

To test our framework in turbulent channel flow, we use direct numerical simulation (DNS) data from \citet{lee2015direct} and from Jim{\'e}nez and co-workers \citep{lozano-duran2014effect, hoyas2006scaling, delalamo2004scaling}. Figure \ref{fig-ch_one_scale}$(a)$ shows the mean velocity gradient profiles at different $Re_\tau=u_\tau\delta/\nu$ from the two datasets. Solid lines correspond to the data from \citet{lee2015direct}, while dashed lines denote those from Jim{\'e}nez and co-workers. We first apply the single-scale self-similarity analysis — that is, the algorithm applied over the full range of $y$ across all stations — where Step 1 identifies the dilation factors $\alpha_j$ and $\gamma_j$ for each station $j$, defined as
\begin{gather}
  \tilde{y}_j = \alpha_jy_j, \quad
  \left.\widetilde{\frac{\mathrm{d}\bar{U}}{\mathrm{d}y}}\right|_j = \gamma_j \left.\frac{\mathrm{d}\bar{U}}{\mathrm{d}y}\right|_j, \label{eq-ch-one-scale-step1}
\end{gather}
where prior knowledge of the problem allows us to set translation factors $\beta_j$ and $\delta_j$ to zero.
Hereafter, the tilde denotes the rescaled variable.
The objective function obtained in Step 1 is 0.046.
Figure \ref{fig-ch_one_scale}$(b)$ shows the rescaled mean velocity gradient profiles obtained using these similarity variables. 
Both the fixed-threshold criterion and visual inspection of the collapse indicate that single-scale self-similarity is insufficient to collapse the profiles over the full $y$ range.
In Step 2, symbolic regression reveals the following expression when $u_\tau$, $\delta$ and $\delta_\nu$ are selected as candidate variables,
\begin{gather}
  \tilde{y} = \delta^{-1.000}\delta_\nu^{-0.000}y, \quad  \widetilde{\frac{\mathrm{d}\bar{U}}{\mathrm{d}y}} = u_\tau^{-1}\delta^{0.992}\delta_\nu^{0.008}\frac{\mathrm{d}\bar{U}}{\mathrm{d}y}, \label{eq-ch-one-scale-1}
\end{gather}
where the exponent of $u_\tau$ is exactly $-1$ due to the dimensional homogeneity constraint. 
Figure \ref{fig-ch_one_scale}$(c)$ presents the rescaled mean velocity gradient profiles obtained using \eqref{eq-ch-one-scale-1}, and the corresponding matched regions for each station are highlighted as shaded areas in figure \ref{fig-ch_one_scale}$(d)$. The profiles are premultiplied by $10^{-j}$ in figure \ref{fig-ch_one_scale}$(d)$ to improve visual separation, with indices $j$ defined in the caption. Since the single-scale analysis is applied over the full $y$ range, the matched ranges of all stations span the full range by construction.
Equation \eqref{eq-ch-one-scale-1} indeed captures the outer similarity, characterised by the outer length scale $\delta$, and successfully collapses the profiles in the outer region, while failing to do so in the near-wall inner region. 
This preferential capture of the outer similarity by the single-scale analysis can be attributed to the uniform linear spacing of the transformed coordinate $\tilde{y}$ grid, where the objective function is evaluated.
The outer similarity then occupies a larger fraction of the grid points in $\tilde{y}$ and therefore dominates the optimisation process.
When the grid points are instead logarithmically spaced, where the inner similarity occupies most of the range, the single-scale analysis identifies the following expression:
\begin{gather}
  \tilde{y} = \delta^{0.021}\delta_\nu^{-1.021}y, \quad  \widetilde{\frac{\mathrm{d}\bar{U}}{\mathrm{d}y}} = u_\tau^{-1}\delta^{0.004}\delta_\nu^{0.996}\frac{\mathrm{d}\bar{U}}{\mathrm{d}y}, \label{eq-ch-one-scale-2}
\end{gather}
which corresponds to the inner similarity characterised by the viscous length scale $\delta_\nu$. In this case, the objective function is 0.008. For brevity, only the final expressions identified by the two-step procedure are presented here and throughout the remainder of the paper, omitting the intermediate results from Step 1.

The recovery of different similarities by the single-scale analysis under different grid distributions, together with the failure of either similarity to collapse the full range of profiles, highlights the multiscale nature of turbulent channel flow. We therefore apply the two-scale self-similarity analysis in the following section.

\begin{figure}
  \centerline{\includegraphics[width=\textwidth]{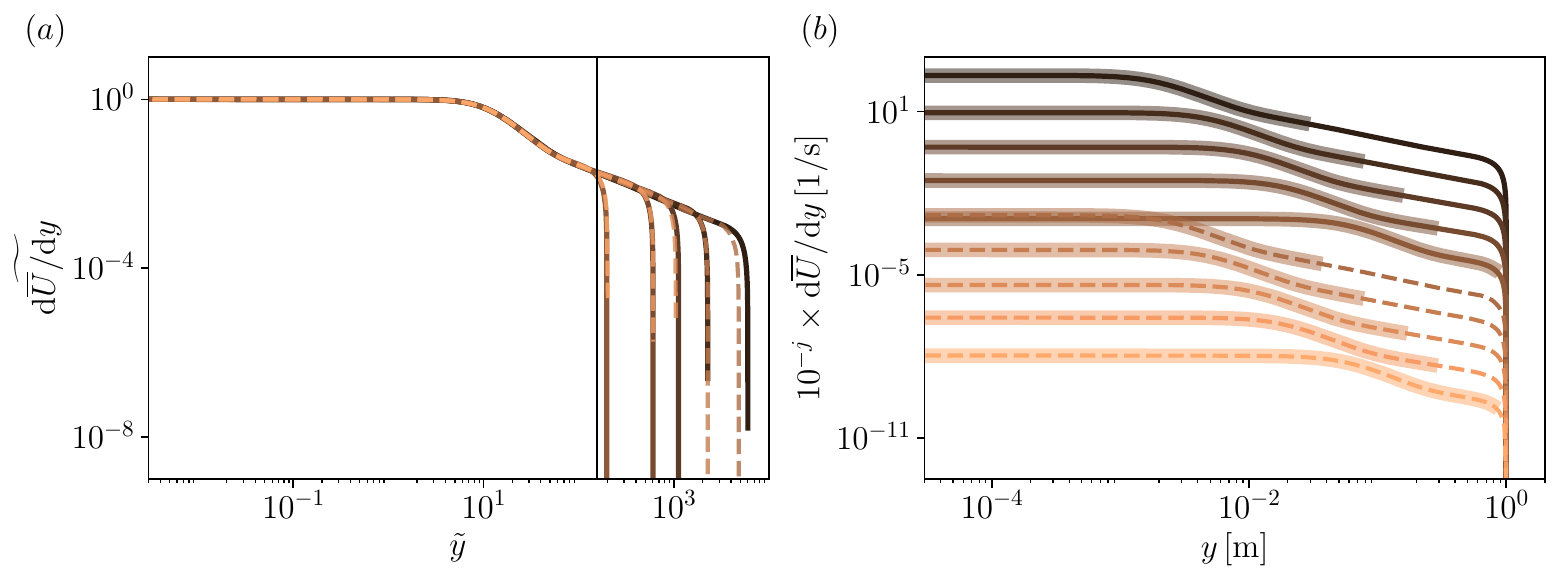}}
  \caption{$(a)$ Rescaled mean velocity gradient profiles using the inner similarity expression \eqref{eq-ch-inner-1}, identified for $\tilde{y} < 162.6$. The vertical line indicates $\tilde{y} = 162.6$. $(b)$ The corresponding collapsed range (shaded) of each profile for the inner similarity. The profiles are premultiplied by $10^{-j}$ for visual separation, where $j$ is the index of the station defined in the caption of figure \ref{fig-ch_one_scale}.}
  \label{fig-ch_inner_1}
\end{figure}

\begin{figure}
  \centerline{\includegraphics[width=\textwidth]{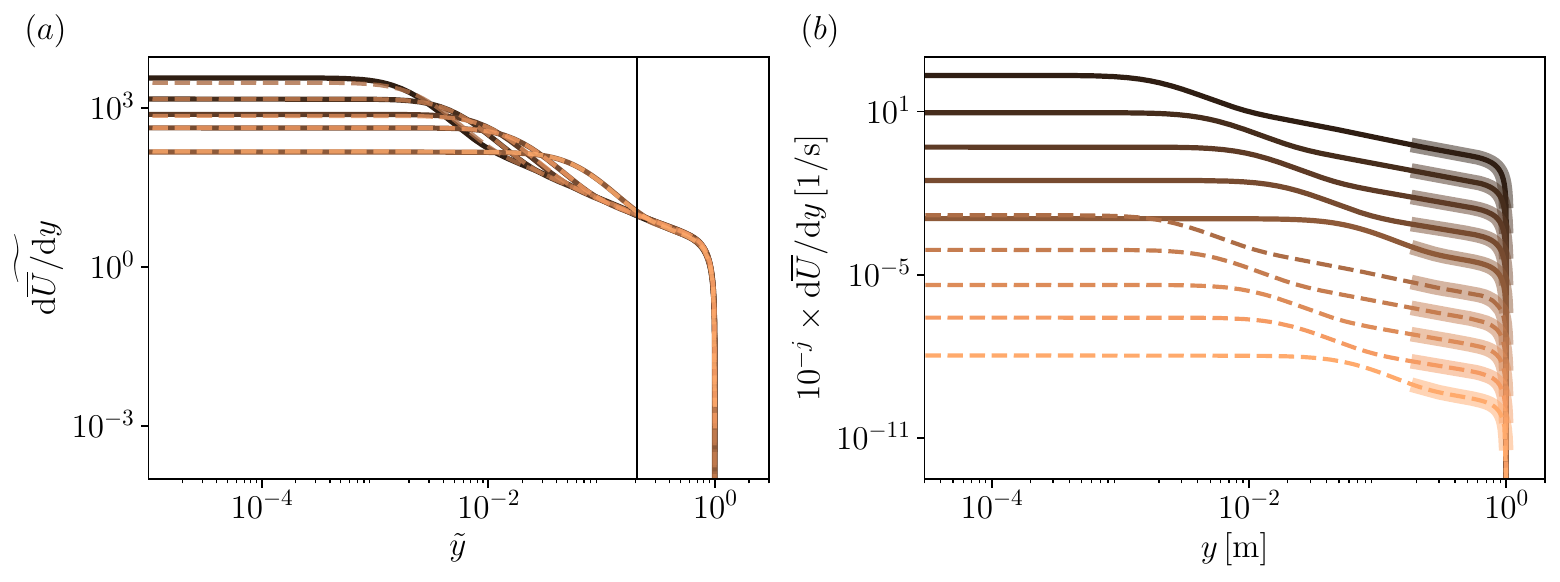}}
  \caption{$(a)$ Rescaled mean velocity gradient profiles using the outer similarity expression \eqref{eq-ch-outer-1}, identified for $\tilde{y} > 0.207$. The vertical line indicates $\tilde{y} = 0.207$. $(b)$ The corresponding collapsed range (shaded) of each profile for the outer similarity. The profiles are premultiplied by $10^{-j}$ for visual separation, where $j$ is the index of the station defined in the caption of figure \ref{fig-ch_one_scale}.}
  \label{fig-ch_outer_1}
\end{figure}

\subsubsection{Two-scale self-similarity} \label{sec-ch-two-scale}

In the two-scale self-similarity analysis, inner and outer similarities are sought by gradually truncating the $y$ range from the opposite end until the computed objective function falls below a prescribed threshold (i.e. 0.001). With the station $Re_\tau=182$ chosen as the reference, the algorithm reveals the existence of inner and outer similarities, given by the following expressions:
\begin{gather}
  \text{Inner:}\quad \tilde{y}_i  = \delta^{0.008}\delta_\nu^{-1.008}y, \quad  \left.\widetilde{\frac{\mathrm{d}\bar{U}}{\mathrm{d}y}}\right|_i = u_\tau^{-1}\delta^{0.001}\delta_\nu^{0.999}\frac{\mathrm{d}\bar{U}}{\mathrm{d}y}, \label{eq-ch-inner-1}\\
  \text{Outer:}\quad \tilde{y}_o = \delta^{-1.000}\delta_\nu^{0.000}y, \quad  \left.\widetilde{\frac{\mathrm{d}\bar{U}}{\mathrm{d}y}}\right|_o = u_\tau^{-1}\delta^{0.962}\delta_\nu^{0.038}\frac{\mathrm{d}\bar{U}}{\mathrm{d}y}. \label{eq-ch-outer-1}
\end{gather}
The sensitivity of the identified expressions to different grid distributions, station sets, and reference stations is examined in Appendix \ref{app-ch-sensitivity}, confirming their robustness. 
Equations \eqref{eq-ch-inner-1} and \eqref{eq-ch-outer-1} are consistent with the traditional understanding of turbulent channel flow, in which the inner region is characterised by the viscous length scale $\delta_\nu$, the outer region by the channel half-height $\delta$, and both scaled by the friction velocity $u_\tau$.
When different libraries of candidate variables are selected, symbolic regression in Step 2 can identify various analytic forms of the similarity variables.
A different velocity scale for the outer region, $U_c-\bar{U}$, proposed by \citet{zagarola1997scaling, zagarola1998mean} based on experimental measurements of turbulent pipe flow, is examined in Appendix \ref{app-ch-outer}. Although pipe-flow velocity profiles scaled by $U_c-\bar{U}$ exhibit better collapse than those scaled by $u_\tau$, our results suggest a similar level of collapse using either of the two velocity scales in turbulent channel flow, consistent with the findings of \citet{pirozzoli2023outer}.
In the following analysis, we therefore use $u_\tau$ as the velocity scale for the outer region without loss of generality.

Figures \ref{fig-ch_inner_1}$(a)$ and \ref{fig-ch_outer_1}$(a)$ demonstrate the successful collapse of the rescaled mean velocity gradient profiles in the inner region using \eqref{eq-ch-inner-1} and in the outer region using \eqref{eq-ch-outer-1}, respectively. 
The inner similarity is identified in the region $\tilde{y}_i<162.6$, and the outer similarity in the region $\tilde{y}_o>0.207$. Both values are indicated by vertical lines in figures \ref{fig-ch_inner_1}$(a)$ and \ref{fig-ch_outer_1}$(a)$. 
The limits of these two regions are determined by the station with the lowest $Re_\tau$, 182, which is the first to deviate from the collapsed curves. 
These two ranges expand if low-$Re_\tau$ stations are excluded from the analysis, as discussed in Appendix \ref{app-ch-reynolds}.
We therefore examine whether an overlap exists between the inner and outer similarity regions at the lowest $Re_\tau$ station, $Re_\tau=182$.
When $\tilde{y}_i$ and $\tilde{y}_o$ are rescaled back to $y$, the corresponding collapsed regions for each station are highlighted as shaded areas in figures \ref{fig-ch_inner_1}$(b)$ and \ref{fig-ch_outer_1}$(b)$. 
For the station $Re_\tau=182$, the inner and outer regions in the $y$ scale correspond to $y<0.855$ and $y>0.207$. The substantial overlap between these two regions indicates two-scale self-similarity in turbulent channel flow and implies an asymptotic matching law in the overlap region. The latter is further investigated in \S\ref{sec-ch-intermediate}.

\begin{table}
\centering
\begin{tabular}{cccccc}
\hline
Index & $y/\delta$ range & $y/\delta_\nu$ range & $a$ & $b$ & $J_m$ \\ \hline
1 & $[0.0010, 0.0057]$ & $[5.3, 29.7]$ & 0.0022 & 0.0049 & 0.0000 \\ \hline
2 & $[0.0014, 0.0080]$ & $[7.4, 41.4]$ & 0.0057 & 0.0092 & 0.0000 \\ \hline
3 & $[0.0020, 0.0112]$ & $[10.5, 57.9]$ & 0.0014 & 0.0015 & 0.0000 \\ \hline
4 & $[0.0029, 0.0159]$ & $[14.8, 82.5]$ & -0.0063 & -0.0113 & 0.0000 \\ \hline
5 & $[0.0040, 0.0223]$ & $[20.7, 115.5]$ & -0.0006 & -0.0034 & 0.0000 \\ \hline
6 & $[0.0057, 0.0316]$ & $[29.7, 163.9]$ & 0.0080 & 0.0078 & 0.0000 \\ \hline
7 & $[0.0080, 0.0449]$ & $[41.4, 232.7]$ & 0.0251 & 0.0274 & 0.0000 \\ \hline
8 & $[0.0112, 0.0631]$ & $[57.9, 327.1]$ & 0.0762 & 0.0792 & 0.0001 \\ \hline
9 & $[0.0159, 0.0891]$ & $[82.5, 462.1]$ & -0.0950 & -0.0648 & 0.0002 \\ \hline
10 & $[0.0223, 0.1261]$ & $[115.5, 653.7]$ & 0.1627 & 0.1830 & 0.0006 \\ \hline
11 & $[0.0316, 0.1778]$ & $[163.9, 922.2]$ & 1.2835 & 1.2166 & 0.0010 \\ \hline
12 & $[0.0449, 0.2518]$ & $[232.7, 1306.0]$ & 0.7651 & 0.7744 & 0.0013 \\ \hline
13 & $[0.0631, 0.3547]$ & $[327.1, 1839.2]$ & 2.1157 & 2.0205 & 0.0009 \\ \hline
14 & $[0.0891, 0.5015]$ & $[462.1, 2601.0]$ & 1.1906 & 1.2022 & 0.0009 \\ \hline
15 & $[0.1261, 0.7080]$ & $[653.7, 3671.7]$ & 1.5562 & 1.5424 & 0.0008 \\ \hline
16 & $[0.1778, 1.0000]$ & $[922.2, 5185.9]$ & 1.0000 & 1.0562 & 0.0009 \\ \hline
\end{tabular} 
\caption{Sliding window ranges for the reference station ($Re_\tau=5186$) with corresponding optimal exponents $a$ and $b$, and the minimum objective function $J_m$ identified by the algorithm.}\label{table-ch}
\end{table}

\begin{figure}
  \centerline{\includegraphics[width=\textwidth]{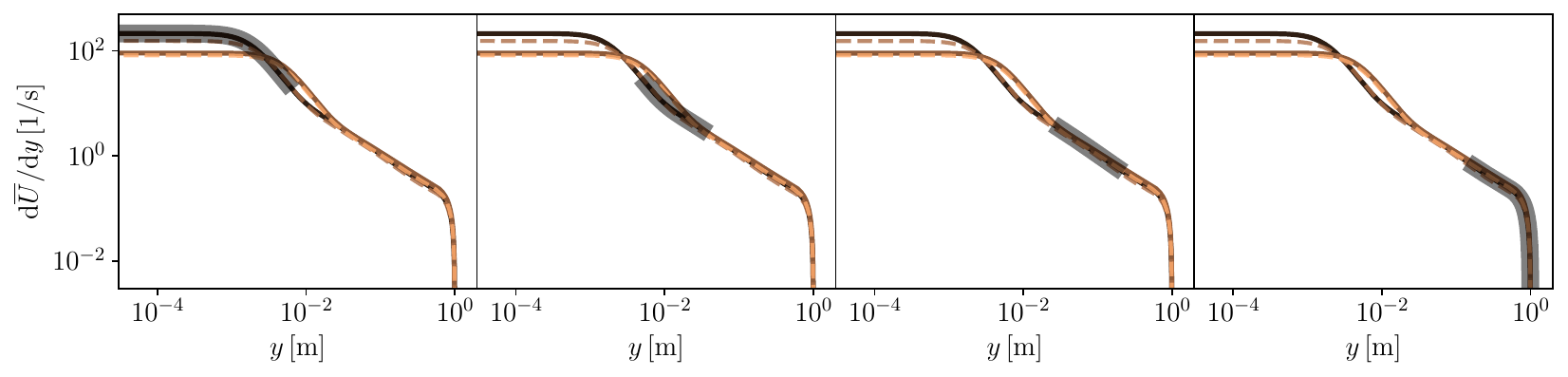}}
  \caption{Sliding window approach for investigating the scaling behaviour in the intermediate region of turbulent channel flow. The ranges covered by windows 1, 6, 11, and 16 on the reference station profile ($Re_\tau = 5186$, black curve) are highlighted by shaded regions, from left to right.}
  \label{fig-sliding_windows}
\end{figure}

\subsubsection{Intermediate region}\label{sec-ch-intermediate}

The intermediate region between the inner and outer layers of turbulent channel flow is of particular importance, as the mean velocity profile within this region is known to follow the logarithmic law, reflecting the scale separation between the two layers. The logarithmic law is expressed as
\begin{gather}
  \frac{\bar{U}}{u_\tau} = \frac{1}{A} \ln \left(\frac{y}{\delta_\nu}\right) + B, \label{eq-log-law}
\end{gather}
where $A$ and $B$ are constants. Equation \eqref{eq-log-law} can be obtained by matching the inner and outer solutions, defined by the similarity variables in \eqref{eq-ch-inner-1} and \eqref{eq-ch-outer-1}, in the overlap region ($\tilde{y}_i \to \infty$, $\tilde{y}_o \to 0$), and then integrating with respect to $y$ \citep{tennekes1972first}. A rigorous derivation was first provided by \citet{millikan1938critical} before the development of the method of matched asymptotic expansions. In what follows, we apply the proposed algorithm to track the transition from inner to outer similarity and, more importantly, to demonstrate a new data-driven approach for recovering the asymptotic matching law, namely the logarithmic law.


\begin{figure}
  \centerline{\includegraphics[width=\textwidth]{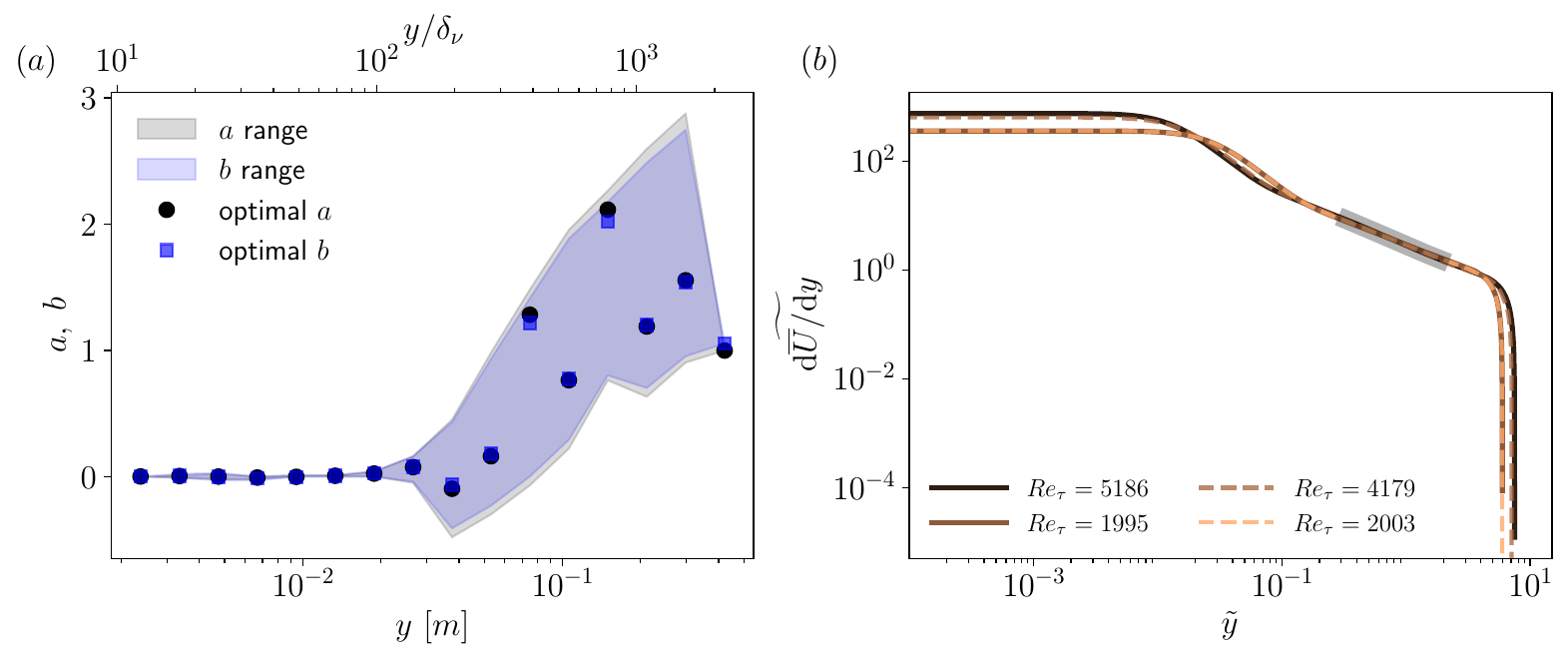}}
  \caption{$(a)$ Variation of the exponents $a$ (black circles) and $b$ (blue squares), as defined in \eqref{eq-ch-intermediate}, with the location of a fixed-length sliding window. The shaded areas highlight the admissible range of $a$ and $b$ values where $J<1.5J_m$ for each window. $(b)$ Rescaled mean velocity gradient profiles using the expression identified for window 12, which has the largest $J_m$ value of 0.0013 among all windows, obtained at $(a,b)=(0.765,0.774)$. The shaded region shows the collapsed range of the window. Four profiles are included in the analysis.}
  \label{fig-ch_ab_y_4}
\end{figure}

To examine how the local scaling transitions from inner to outer regions, we use a sliding window approach in which a fixed-length window is defined on the logarithmic $y$ scale of the reference station and slid across the full $y$ range, as shown in figure \ref{fig-sliding_windows}. The four profiles with the highest $Re_\tau$ are selected for analysis, as they possess a sufficiently large intermediate region, with the profile at $Re_\tau = 5186$ serving as the reference. For each window position, the algorithm identifies the matching ranges in the other stations and determines the corresponding self-similarity expressions.
To facilitate the robustness in the tracking of the transition of the self-similarity expressions, we incorporate the inner and outer expressions \eqref{eq-ch-inner-1} and \eqref{eq-ch-outer-1} from the preceding analysis and search for the optimal exponents $a$ and $b$ in the following variables: 
\begin{gather}
  \tilde{y} = \frac{y}{\delta^{a}\delta_\nu^{1-a}}, \quad  \widetilde{\frac{\mathrm{d}\bar{U}}{\mathrm{d}y}} = \frac{\delta^{b}\delta_\nu^{1-b}}{u_\tau}\frac{\mathrm{d}\bar{U}}{\mathrm{d}y}. \label{eq-ch-intermediate}
\end{gather}
Here, the length scales associated with $y$ and $\mathrm{d}\bar{U}/\mathrm{d}y$ are expressed in terms of both the inner and outer scales, $\delta_\nu$ and $\delta$.
The variable $\tilde{y}$ in \eqref{eq-ch-intermediate} is also known as the intermediate variable in the method of matched asymptotic expansions, where $a$ is typically constrained to $[0,1]$ \citep{bender1999advanced}. We show below that for this multiscale problem, where the inner and outer solutions share the same monomial form and differ only in their characteristic length scales, no such restriction is required and $a$ and $b$ may take arbitrary values.

With the window length set to $0.75$ decades (one fourth of the range from 0.001 to 1 in logarithmic scale) and an 80\% overlap between successive windows, table \ref{table-ch} lists the $y$ range of each window together with the corresponding optimal values of $a$ and $b$ and the minimum objective function $J_m$.
The algorithm reveals that the exponents $a$ and $b$, shown by black circles and blue squares in figure \ref{fig-ch_ab_y_4}$(a)$, respectively, are in close agreement across the entire $y$ range, suggesting self-similarity in the form of \eqref{eq-ch-intermediate} for each window.
In the inner and outer regions, both $a$ and $b$ are close to 0 and 1, respectively, consistent with expressions \eqref{eq-ch-inner-1} and \eqref{eq-ch-outer-1}. However, in the intermediate region, the values fall outside $[0,1]$ while still providing a good collapse of the profiles, as shown in figure \ref{fig-ch_ab_y_4}$(b)$ for window 12, which has the largest $J_m$ value of 0.0013 among all windows. We therefore further explore this region using sampling methods to find the range of $a$ and $b$ that can provide a sufficiently good collapse of the profiles, defined as $(a, b)$ satisfying $J < \omega J_m$ for a prescribed multiplier $\omega > 1$.
The shaded region in figure \ref{fig-ch_ab_y_4}$(a)$ shows the admissible ranges when $\omega = 1.5$.
The admissible ranges are substantially wider in the intermediate region than in the inner and outer regions, indicating that the intermediate region is less sensitive to the precise choice of $(a, b)$ and that a reference window there achieves good collapse with matched windows over a wider range of positions in the other stations.
These admissible ranges become narrower in the intermediate region when two lower-$Re_\tau$ stations are included in the analysis, as shown in figure \ref{fig-ch_ab_y_6}$(a)$ in Appendix \ref{app-intermediate-fig}, and even narrower when all ten stations are included (figure \ref{fig-ch_ab_y_10}$(a)$).
This is because the lower-$Re_\tau$ stations have a smaller intermediate region, which limits the available matched windows and thereby reduces the admissible ranges of $(a, b)$.

\begin{figure}
  \centerline{\includegraphics[width=0.5\textwidth]{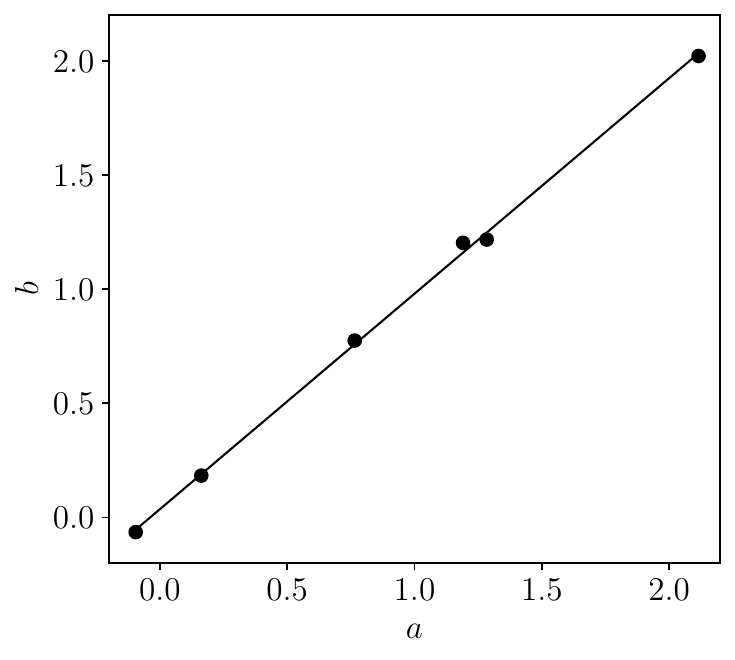}}
  \caption{Relationship between the exponents $a$ and $b$ in the intermediate range. The optimal values of $a$ and $b$ corresponding to windows 9 to 15 in figure \ref{fig-ch_ab_y_4} are employed in a linear fit, which gives $b = 0.9436a + 0.0356$ (black line).}
  \label{fig-ch_a_b_fit}
\end{figure}

For velocity profiles at asymptotically large $Re_\tau$, the data-driven results above indicate the existence of an intermediate region in which the profiles collapse under essentially arbitrary choices of $a$ and $b$, with $a \approx b$. This suggests that the mean-flow structure in this region is not governed by any intrinsic or distinguished characteristic length scale, consistent with the scale-free nature of the logarithmic region in turbulent channel flow. 
More importantly, this scaling behaviour identified directly from the data by the algorithm leads to the classical logarithmic law, providing a data-driven route to its derivation. A mathematical derivation is given below.

We start from the following similarity expression discovered by the algorithm in the intermediate range,
\begin{gather}
  \frac{\mathrm{d}\bar{U}}{\mathrm{d}y} = \frac{u_\tau}{\delta^{b}\delta_\nu^{1-b}}f\left(\frac{y}{\delta^{a}\delta_\nu^{1-a}}\right),\quad a\approx b, \label{eq-ch-overlap}
\end{gather}
where $f$ is a nondimensional function, and $a, b$ can take arbitrary values. We consider two limiting cases: $a_0=0, b_0\approx0$ and $a_1=1, b_1\approx1$. Substituting the two limits of $a$ into \eqref{eq-ch-overlap} yields
\begin{align}
  \frac{\mathrm{d}\bar{U}}{\mathrm{d}y} &= \frac{u_\tau}{\delta^{b_0}\delta_\nu^{1-b_0}}f_0\left(\frac{y}{\delta_\nu}\right),\label{eq-ch-a0b0}\\
  \frac{\mathrm{d}\bar{U}}{\mathrm{d}y} &= \frac{u_\tau}{\delta^{b_1}\delta_\nu^{1-b_1}}f_1\left(\frac{y}{\delta}\right). \label{eq-ch-a1b1}
\end{align}
Equating them then leads to
\begin{gather}
  \frac{(y/\delta_\nu)^{b_1}}{(y/\delta_\nu)^{b_0}}f_0\left(\frac{y}{\delta_\nu}\right) = \frac{(y/\delta)^{b_1}}{(y/\delta)^{b_0}}f_1\left(\frac{y}{\delta}\right).
\end{gather}
Since the left- and right-hand sides depend only on inner and outer variables, $y/\delta_\nu$ and $y/\delta$, respectively, they must be equal to a constant. We thus obtain 
\begin{gather}
  f_0(Y) = f_1(Y) \propto Y^{b_0-b_1}, \label{eq-ch-f0f1}
\end{gather}
where $Y$ is the argument of the function. Substituting \eqref{eq-ch-f0f1} back into \eqref{eq-ch-a0b0} or \eqref{eq-ch-a1b1} then gives
\begin{gather}
  \frac{\mathrm{d}\bar{U}}{\mathrm{d}y} \propto \frac{u_\tau}{y}\left(\frac{y}{\delta}\right)^{b_0}\left(\frac{y}{\delta_\nu}\right)^{1-b_1}, \label{eq-ch-overlap-1}
\end{gather}
where $b_0$ and $b_1$ can be determined from the relation between $a$ and $b$, obtained from the linear fit of the computed values that yield a good collapse in the intermediate range. Here, we choose the optimal values of $a$ and $b$ corresponding to windows 9 to 15 and employ them in a linear fit, as shown in figure \ref{fig-ch_a_b_fit}, which gives
\begin{gather}
  b = 0.9436a + 0.0356.
\end{gather}
Equation \eqref{eq-ch-overlap-1} then becomes
\begin{gather}
  \frac{\mathrm{d}\bar{U}}{\mathrm{d}y} \propto \frac{u_\tau}{y}\left(\frac{y}{\delta}\right)^{0.036}\left(\frac{y}{\delta_\nu}\right)^{0.021}. \label{eq-ch-overlap-2}
\end{gather}
Equation \eqref{eq-ch-overlap-2} shows that the data-driven scaling recovers the logarithmic law as the leading-order behaviour. If the asymptotically small exponent corrections associated with $y/\delta$ and $y/\delta_\nu$ are neglected, the equation reduces to $\mathrm{d}\bar{U}/\mathrm{d}y \propto u_\tau/y$, and integration with respect to $y$ gives the classical logarithmic law \eqref{eq-log-law}. 

Rewriting \eqref{eq-ch-overlap-2} in terms of $Re_\tau$ yields
\begin{gather}
  \frac{\mathrm{d}\bar{U}}{\mathrm{d}y} \propto \frac{u_\tau}{y}\left(\frac{1}{Re_\tau}\right)^{0.036}\left(\frac{y}{\delta_\nu}\right)^{0.057},
\end{gather}
which retains a finite-$Re_\tau$ correction and has the same structure as the power-law scaling proposed by \citet{barenblatt1993scalinga} and \citet{barenblatt1993scalingb} as an alternative to the logarithmic law:
\begin{gather}
  \frac{\mathrm{d}\bar{U}}{\mathrm{d}y} = \frac{u_\tau}{y}\Phi\left(\frac{y}{\delta_\nu},Re\right), \quad \text{with} \quad \Phi\left(\frac{y}{\delta_\nu},Re\right) = C\left(\frac{y}{\delta_\nu}\right)^m, \label{eq-ch-barenblatt}
\end{gather}
where $C$ and $m$ are Reynolds-number-dependent parameters. Our calculation yields $C \propto Re_\tau^{-0.036}$ and $m \approx 0.057$.

This demonstrates a key strength of the algorithm: it recovers the leading-order logarithmic scaling while also quantifying the finite $Re_\tau$ power-law correction.

\subsection{Late-stage homogeneous decaying turbulence} \label{sec-decaying_turbulence}

We then consider the last problem left open in \citet{bempedelis2025extracting}: the self-similarity of homogeneous decaying turbulence. Decaying turbulence, which can be generated experimentally by passing a uniform stream through a grid, and numerically by removing the forcing from turbulence in a periodic box, provides a good approximation to homogeneous isotropic turbulence and is therefore well suited for the analysis of self-similarity and scaling laws in turbulent energy cascade \citep{ishihara2009study}.
The classical theory of Kolmogorov describes an equilibrium energy cascade in high-Reynolds-number turbulence, in which kinetic energy is transferred from large to progressively smaller scales, where it is ultimately dissipated as heat. Under certain conditions, this cascade is known to lead to the classical dissipation scaling $\epsilon = C_\epsilon u^3/L$ with $C_\epsilon = \text{const.}$, where $\epsilon$ is the dissipation rate, $u = \sqrt{2K/3}$ denotes the characteristic velocity scale based on the turbulent kinetic energy $K$, and $L$ is the integral length scale. Numerous experimental and numerical studies have provided strong support for this classical dissipation scaling in the late stage of homogeneous decaying turbulence, while a different dissipation scaling emerges in the early stage \citep{vassilicos2015dissipation}.
In this section, we focus on late-stage homogeneous decaying turbulence, where the classical dissipation scaling holds. The self-similarity of early-stage homogeneous decaying turbulence, characterised by nonclassical dissipation scaling, is then investigated in \S\ref{sec-decaying_turbulence_ic}.

At sufficiently high Reynolds numbers, classical Kolmogorov theory predicts a universal form for the statistics of small-scale turbulent motion, uniquely determined by the dissipation rate $\epsilon$ and the kinematic viscosity $\nu$ \citep{kolmogorov1941dissipation,kolmogorov1941local},
\begin{gather}
  E(\kappa) = \epsilon^{2/3}\kappa^{-5/3}\Psi(\kappa\eta), \label{eq-K41-a}
\end{gather}
where $E$ is the energy spectrum, $\kappa$ is the wavenumber, $\Psi(\kappa\eta)$ is a universal nondimensional function, 
and $\eta = (\nu^3/\epsilon)^{1/4}$ is the Kolmogorov length scale. In the limit $\kappa\eta\to0$, Kolmogorov's theory further predicts that the statistics are independent of the viscosity $\nu$, implying that $\Psi(\kappa\eta)$ approaches a constant. In this regime, the universal energy spectrum relation \eqref{eq-K41-a} becomes
\begin{gather}
  E(\kappa) \propto\epsilon^{2/3}\kappa^{-5/3}, \label{eq-53law}
\end{gather}
which is the well-known $-5/3$ law governing the so-called inertial subrange.

Kolmogorov's theory provides a clear description of the energy spectrum for small-scale motions. However, it does not account for the spectrum associated with large-scale motions. A model spectrum that describes the energy distribution across all wavenumbers is proposed by \citet[p.~232]{pope2000turbulent}:
\begin{gather}
  E(\kappa) \propto \epsilon^{2/3}\kappa^{-5/3}f_L(\kappa L)f_\eta(\kappa\eta), \label{eq-model-spectrum}
\end{gather}
where $f_L(\kappa L)$ and $f_\eta(\kappa\eta)$ are specified nondimensional functions that determine deviations from the $-5/3$ law in the large- and small-scale motions, respectively. Both $f_L(\kappa L)$ and $f_\eta(\kappa\eta)$ approach a constant in the inertial subrange, i.e. $\kappa L\to\infty$ and $\kappa\eta\to0$, and the $-5/3$ law \eqref{eq-53law} is recovered. The above model suggests a two-scale nature of the energy spectrum, with the integral length scale $L$ and the Kolmogorov length scale $\eta$ serving as the characteristic length scales for the large- and small-scale motions, respectively.
The model spectrum \eqref{eq-model-spectrum} was extended by \citet{meyers2008functional} to include a predissipative bottleneck effect and an intermittency correction.
The same characteristic length scales were also derived by \citet{lundgren2002kolmogorov,lundgren2003kolmogorov}, who rigorously analysed Kolmogorov's theory using matched asymptotic expansions. In this section, we apply the data-driven self-similarity framework to the energy spectrum of decaying turbulence, following the same two-scale perspective.  

\begin{figure}
  \centerline{\includegraphics[width=\textwidth]{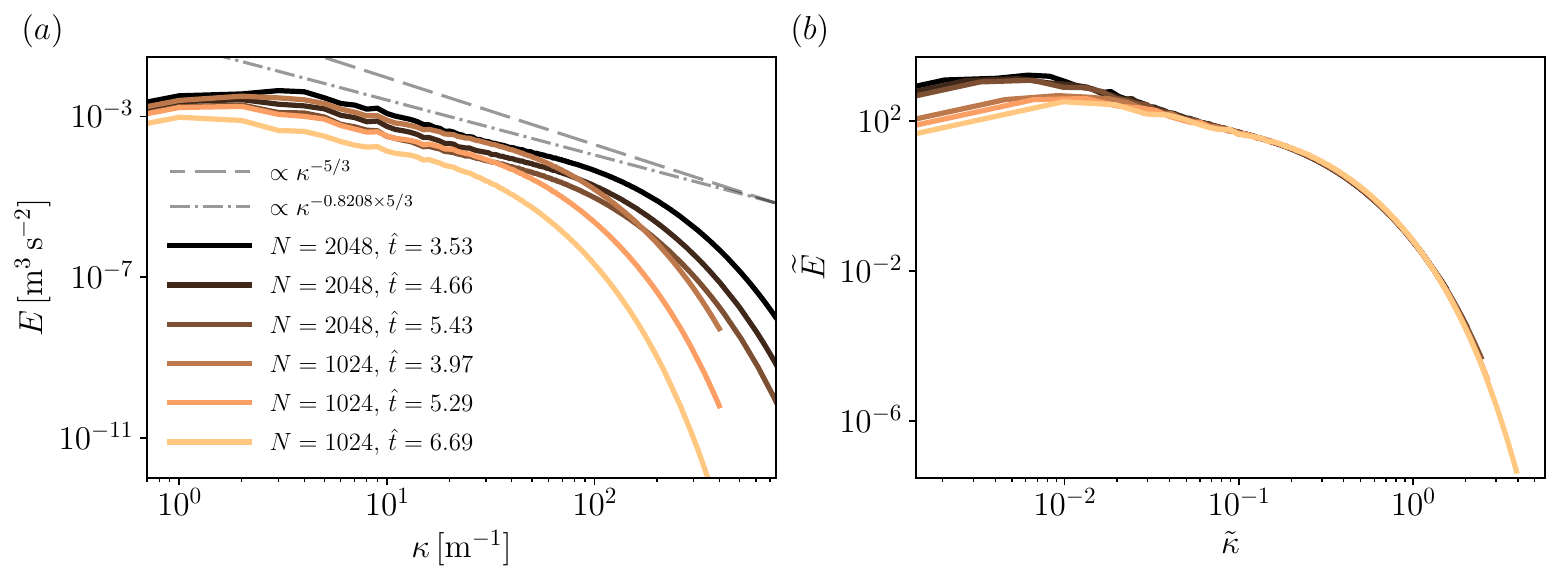}}
  \caption{$(a)$ Energy spectra $E(\kappa)$ at different simulation sizes $N$ and decay times $\hat{t}$, taken from DNS data of \citet{goto2016unsteady}. The dashed and dash-dotted lines represent the relations $E\propto\kappa^{-5/3}$ and $E\propto\kappa^{-0.8208\times5/3}$, respectively, which will be discussed later in the paper. $(b)$ Rescaled energy spectra $\tilde{E}(\tilde{\kappa})$ using expression \eqref{eq-dt-one-scale-1}, identified through one-scale self-similarity analysis. The same line styles are used in figures \ref{fig-dt_inner}, \ref{fig-dt_outer}--\ref{fig-dt_ab_y_4}, \ref{fig-dt_ab_y_2}--\ref{fig-dt_ab_y_3}.}
  \label{fig-dt_one_scale}
\end{figure}

\subsubsection{Single-scale self-similarity}

We use DNS data for decaying turbulence in a periodic box from \citet{goto2016unsteady}, employing the two largest simulation sizes, $N=1024$ and $N=2048$. Higher values of $N$ correspond to higher Reynolds numbers. 
The stations are chosen at sufficiently large decay times $\hat{t}$, defined as $\hat{t}=\int_0^tu/Ldt$, where the classical dissipation scaling, $C_\epsilon = \epsilon L/u^3 = \text{const.}$, is satisfied. Figure \ref{fig-dt_one_scale}$(a)$ shows the energy spectra $E(\kappa)$ at six different $\hat{t}$ taken from the two simulation sizes, with $\kappa$ ranging from 0 to 400 for case $N=1024$ and from 0 to 750 for case $N=2048$. The single-scale self-similarity framework is first applied. 
When the grid points are uniformly spaced in the transformed coordinate $\tilde{\kappa}$, using linear and logarithmic spacing respectively, the following expressions are obtained:
\begin{gather}
  \tilde{\kappa} = L^{0.030}\eta^{0.970}\kappa, \quad \tilde{E} = \epsilon^{-0.667}(L^{0.086}\eta^{0.914})^{-5/3}E. \label{eq-dt-one-scale-1}\\
  \tilde{\kappa} = L^{0.238}\eta^{0.762}\kappa, \quad \tilde{E} = \epsilon^{-0.667}(L^{0.622}\eta^{0.378})^{-5/3}E. \label{eq-dt-one-scale-2}
\end{gather}
The objective function values are 0.014 and 0.130, respectively.
The translation functions $\beta$ and $\delta$ defined in \eqref{eq-xi} and \eqref{eq-qtilde} are set to zero in Step 1 based on the prior knowledge of the problem.
The symbolic regression in Step 2 is performed using $\{L, \eta\}$ and $\{L, \eta, \epsilon, K\}$ as candidate variable sets to interpret the transformation functions.
In the above equation and in \eqref{eq-dt-inner}--\eqref{eq-dt-outer}, the dissipation scaling $\epsilon\propto K^{3/2}/L$ has been applied to the algorithm outputs to aid interpretation and comparison.

Both the fixed-threshold criterion and visual inspection of the collapse indicate that single-scale self-similarity is insufficient to collapse the energy spectra across the full wavenumber range.
Figure \ref{fig-dt_one_scale}$(b)$ shows the rescaled energy spectra $\tilde{E}$ obtained using expression \eqref{eq-dt-one-scale-1}, which captures the inner similarity because the high-wavenumber region occupies a wider range on the linear scale. By contrast, expression \eqref{eq-dt-one-scale-2} does not correspond to a physically meaningful similarity. Unlike in turbulent channel flow, where the single-scale analysis captures the two similarities under different grid distributions, the outer similarity is not recovered here. This difference is attributed to the grid distributions of the two datasets: the turbulent channel flow data are refined near small $y$, while the decaying-turbulence data are obtained on a uniform grid throughout the $\kappa$ domain.
It is worth noting that the exponents identified in \eqref{eq-dt-one-scale-1}--\eqref{eq-dt-one-scale-2} differ from those reported in equation (3.11) of \citet{bempedelis2025extracting}. This difference arises from the inclusion of additional high-wavenumber data and from a different selection of decay times. The algorithm of \citet{bempedelis2025extracting} requires different station data to share the same range, whereas the present algorithm can accommodate station data with different ranges. This allows the present analysis to include more high-wavenumber data points from the $N=2048$ case.

The dependence of the single-scale results on the grid distribution and their inability to collapse the full range of spectra suggest the multiscale nature of decaying turbulence. We therefore apply the two-scale self-similarity analysis in the following section.

\begin{figure}
  \centerline{\includegraphics[width=\textwidth]{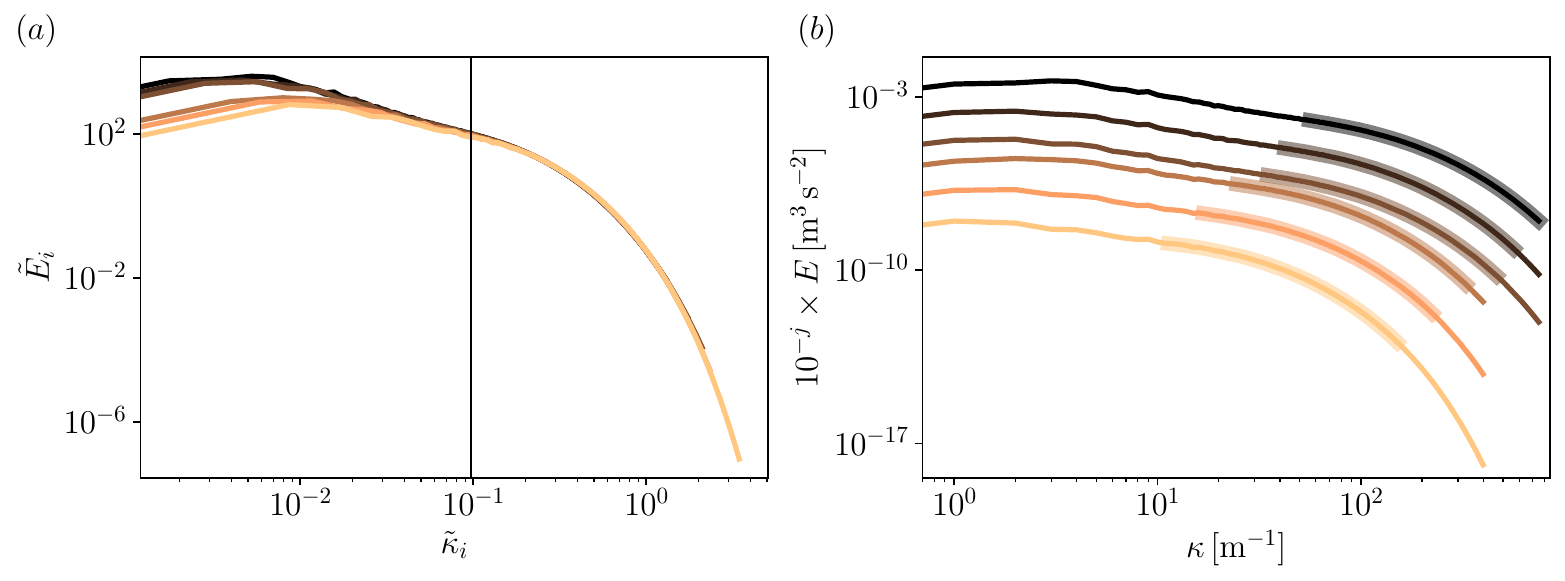}}
  \caption{$(a)$ Rescaled energy spectra $\tilde{E}_i(\tilde{\kappa}_i)$ using the inner similarity expression \eqref{eq-dt-inner}, identified for $\tilde{\kappa}_i>0.098$. The vertical line indicates $\tilde{\kappa}_i=0.098$. $(b)$ The corresponding collapsed range (shaded) of each profile for the inner similarity. The profiles are premultiplied by $10^{-j}$ for visual separation, where $j$ denotes the station index. Indices $j = 0$ to $2$ and $3$ to $5$ correspond to cases $N=2048$ and $N=1024$, respectively, with each group ordered by increasing $\hat{t}$.}
  \label{fig-dt_inner}
\end{figure}

\subsubsection{Two-scale self-similarity} \label{sec-dt-two-scale}

The two-scale self-similarity framework is first applied to search for the inner similarity. With the profile at $N = 2048$ and $\hat{t} = 3.53$ selected as the reference, and with the fixed-threshold criterion applied, the inner similarity is identified as
\begin{gather}
   \text{Inner:}\quad \tilde{\kappa}_i = L^{0.001}\eta^{0.999}\kappa, \quad \tilde{E}_i = \epsilon^{-0.667}(L^{-0.010}\eta^{1.010})^{-5/3}E, \label{eq-dt-inner}
\end{gather}
which is essentially \eqref{eq-K41-a}.
Excellent collapse of the rescaled energy spectra is obtained in the inner region, as shown in figure \ref{fig-dt_inner}$(a)$.

\begin{figure}
  \centerline{\includegraphics[width=\textwidth]{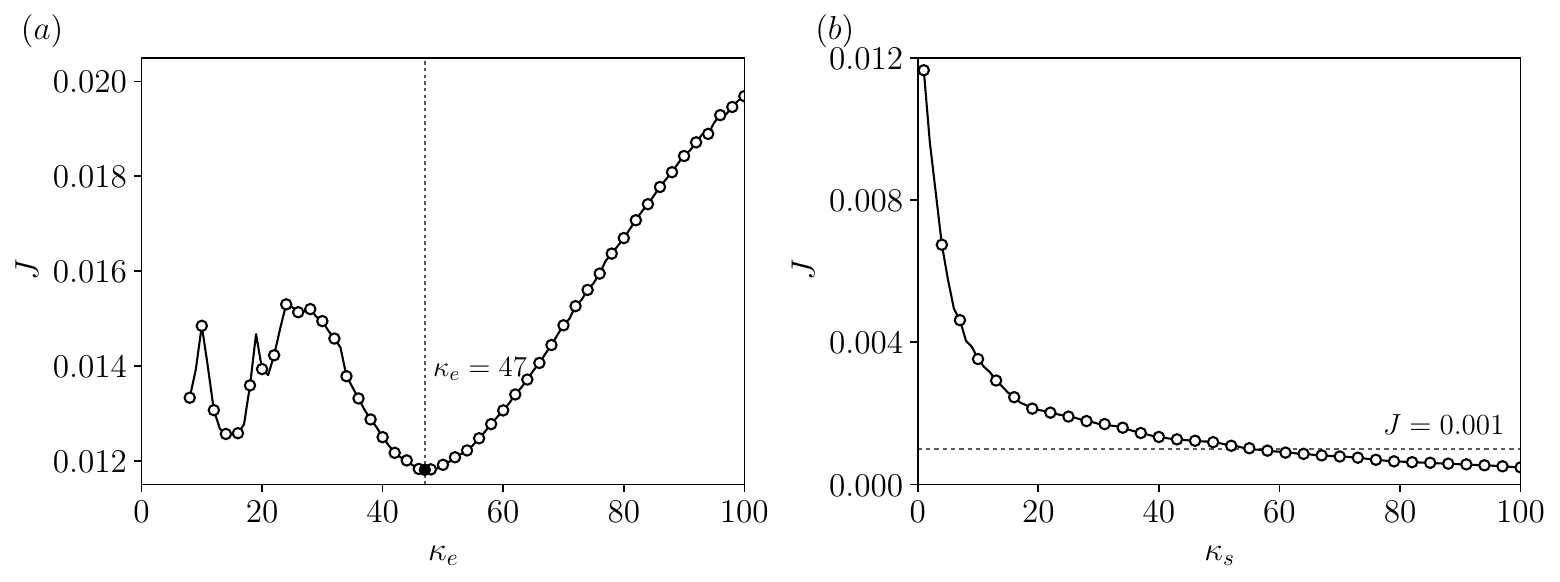}}
  \caption{$(a)$ Variation of the objective function $J$ with the upper limit $\kappa_e$ of the search range $[1,\kappa_e]$ for the outer similarity. The vertical line indicates $\kappa_e=47$. $(b)$ Variation of the objective function $J$ with the lower limit $\kappa_s$ of the search range $[\kappa_s,750]$ for the inner similarity. The horizontal line indicates the threshold of 0.001.}
  \label{fig-dt_loss}
\end{figure}

\begin{figure}
  \centerline{\includegraphics[width=\textwidth]{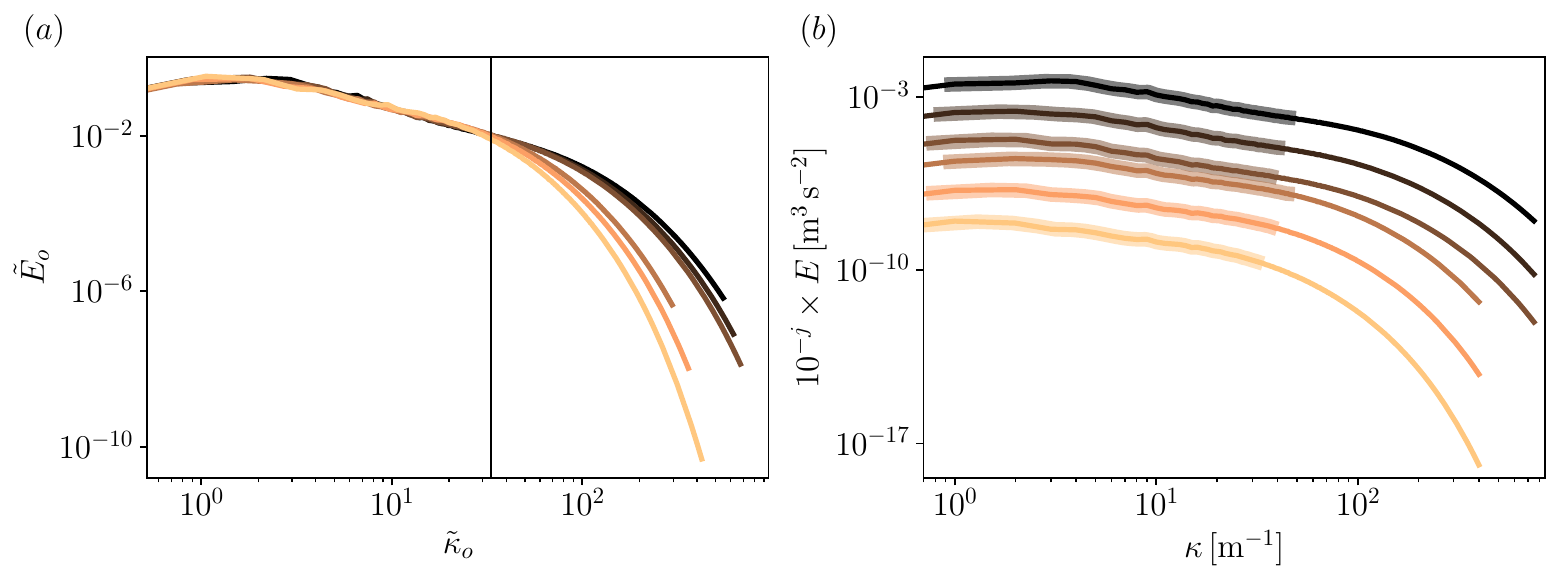}}
  \caption{$(a)$ Rescaled energy spectra $\tilde{E}_o(\tilde{\kappa}_o)$ using the outer similarity expression \eqref{eq-dt-outer}, identified for $\tilde{\kappa}_o<33.24$. The vertical line indicates $\tilde{\kappa}_o = 33.24$. $(b)$ The corresponding collapsed range (shaded) of each profile for the outer similarity. The profiles are premultiplied by $10^{-j}$ for visual separation, where the station index $j$ is defined in the caption of figure \ref{fig-dt_inner}.}
  \label{fig-dt_outer}
\end{figure}

The outer similarity is identified using the global minimum criterion, since the objective function does not fall below the prescribed threshold.
Figure \ref{fig-dt_loss}$(a)$ shows the variation of the objective function $J$ with the upper limit $\kappa_e$ of the search range $[1,\kappa_e]$ for the outer similarity.
For the upper limit $\kappa_e<100$, the objective function $J$ decreases with decreasing $\kappa_e$ and reaches a global minimum value of 0.0118 at $\kappa_e=47$ before oscillating at smaller $\kappa_e$. This behaviour differs from that observed in the inner similarity search (figure \ref{fig-dt_loss}$(b)$) and in the inner and outer similarity searches in turbulent channel flow (not shown), where the objective function decreases monotonically and reaches a plateau as the search range is reduced.
The non-monotonic behaviour of the objective function in the outer region of decaying turbulence may be attributed to several factors. First, compared with the inner region, which spans several hundred wavenumbers, the outer region covers only a narrow range and is therefore more sensitive to noise and numerical error. Second, the calculation of the energy spectrum $E$ at small wavenumbers is prone to error because only a limited number of Fourier modes are available \citep{stepanov2014systematic}; this error also affects the calculation of the integral length scale $L$. Third, after the rapid evolution during the early decay, the energy spectra at late decay times become less distinguishable from one another \citep{steiros2022turbulence,goto2016unsteady}, making it more difficult for the algorithm to identify the intrinsic similarity.
It is therefore expected that the non-monotonic behaviour of the objective function at low wavenumbers would be attenuated with increasing dataset resolution.
The outer similarity expression obtained for the range $[1,47]$ is
\begin{gather}
   \text{Outer:}\quad \tilde{\kappa}_o = L^{1.085}\eta^{-0.085}\kappa, \quad \tilde{E}_o = \epsilon^{-0.667}(L^{1.013}\eta^{-0.013})^{-5/3}E. \label{eq-dt-outer}
\end{gather}
Figure \ref{fig-dt_outer}$(a)$ shows the collapse obtained for the outer region using this expression. Despite the relatively large objective function, the identified outer similarity still achieves a reasonable collapse.

The inner and outer similarities, \eqref{eq-dt-inner} and \eqref{eq-dt-outer}, are identified in the scaled ranges $\tilde{\kappa}_i>0.098$ and $\tilde{\kappa}_o<33.24$, with both values indicated by vertical lines in figures \ref{fig-dt_inner}$(a)$ and \ref{fig-dt_outer}$(a)$, respectively. When rescaled back to $\kappa$, the corresponding collapsed ranges for each station are presented as shaded areas in figures \ref{fig-dt_inner}$(b)$ and \ref{fig-dt_outer}$(b)$. The profiles are premultiplied by $10^{-j}$ for visual separation with indices $j$ defined in the caption of figure \ref{fig-dt_inner}.
The existence of an overlap is examined at the station $N=1024$, $\hat{t}=6.69$, which first deviates from the collapsed curve. Its inner and outer regions in the $\kappa$ scale correspond to $\kappa>11.29$ and $\kappa<31.33$. The considerable overlap of the two regions suggests two-scale self-similarity in late-stage homogeneous decaying turbulence.

\subsubsection{Intermediate region}

To investigate the scaling in the intermediate region, we choose the station $N=2048$, $\hat{t}=3.53$, which exhibits the largest intermediate region, as the reference. 
Following \S\ref{sec-ch-intermediate}, we use the sliding window approach to track the transition in similarity expressions from inner to outer regions. To facilitate the robustness of the tracking, we incorporate the inner and outer expressions from the preceding analysis and search for the optimal exponents $c$ and $d$ in the following similarity variables:
\begin{gather}
    \tilde{\kappa} = L^c\eta^{1-c}\kappa, \quad \tilde{E} = \epsilon^{-2/3}(L^d\eta^{1-d})^{-5/3}E. \label{eq-dt-intermediate}
\end{gather}
A sliding window of the reference station with a length of 0.71 decades (one fourth of the range from 1 to 750 in logarithmic space) is used, with an 80\% overlap between successive windows. The algorithm then identifies the matched ranges from the other stations; the corresponding optimal values of $c$ and $d$ and the minimum objective function $J_m$ for each window are listed in table \ref{table-dt}.

\begin{table}
\centering
\begin{tabular}{ccccc@{\hspace{1.0cm}}ccccc}
\hline
Index & $\kappa$ range & $c$     & $d$     & $J_m$ &
Index & $\kappa$ range & $c$     & $d$     & $J_m$ \\ \hline
1     & $[1, 5]$       & 1.0308  & 1.0457  & 0.0166 &
9     & $[14, 74]$     & 0.4670  & 0.4667  & 0.0031        \\ \hline
2     & $[1, 7]$       & 1.0315  & 1.0598  & 0.0145 &
10    & $[20, 103]$    & 0.2530  & 0.2913  & 0.0017        \\ \hline
3     & $[2, 10]$      & 1.0616  & 1.0716  & 0.0136 &
11    & $[27, 143]$    & 0.1595  & 0.2068  & 0.0010        \\ \hline
4     & $[3, 14]$      & 1.9872  & 1.7120  & 0.0094 &
12    & $[38, 200]$    & 0.1031  & 0.1448  & 0.0005        \\ \hline
5     & $[4, 20]$      & 1.6917  & 1.4763  & 0.0098 &
13    & $[53, 278]$    & 0.0694  & 0.1004  & 0.0003        \\ \hline
6     & $[5, 27]$      & 1.3765  & 1.2076  & 0.0106 &
14    & $[74, 387]$    & 0.0392  & 0.0495  & 0.0002        \\ \hline
7     & $[7, 38]$      & 1.0737  & 0.9541  & 0.0093 &
15    & $[103, 539]$   & 0.0127  & -0.0064 & 0.0002        \\ \hline
8     & $[10, 53]$     & 0.6127  & 0.5828  & 0.0061 &
16    & $[143, 749]$   & -0.0125 & -0.0726 & 0.0003        \\ \hline
\end{tabular} 
\caption{Sliding window ranges for the reference station ($N=2048$, $\hat{t}=3.53$) with corresponding optimal exponents $c$ and $d$, and the minimum objective function $J_m$ identified by the algorithm.}
\label{table-dt}
\end{table}

\begin{figure}
  \centerline{\includegraphics[width=\textwidth]{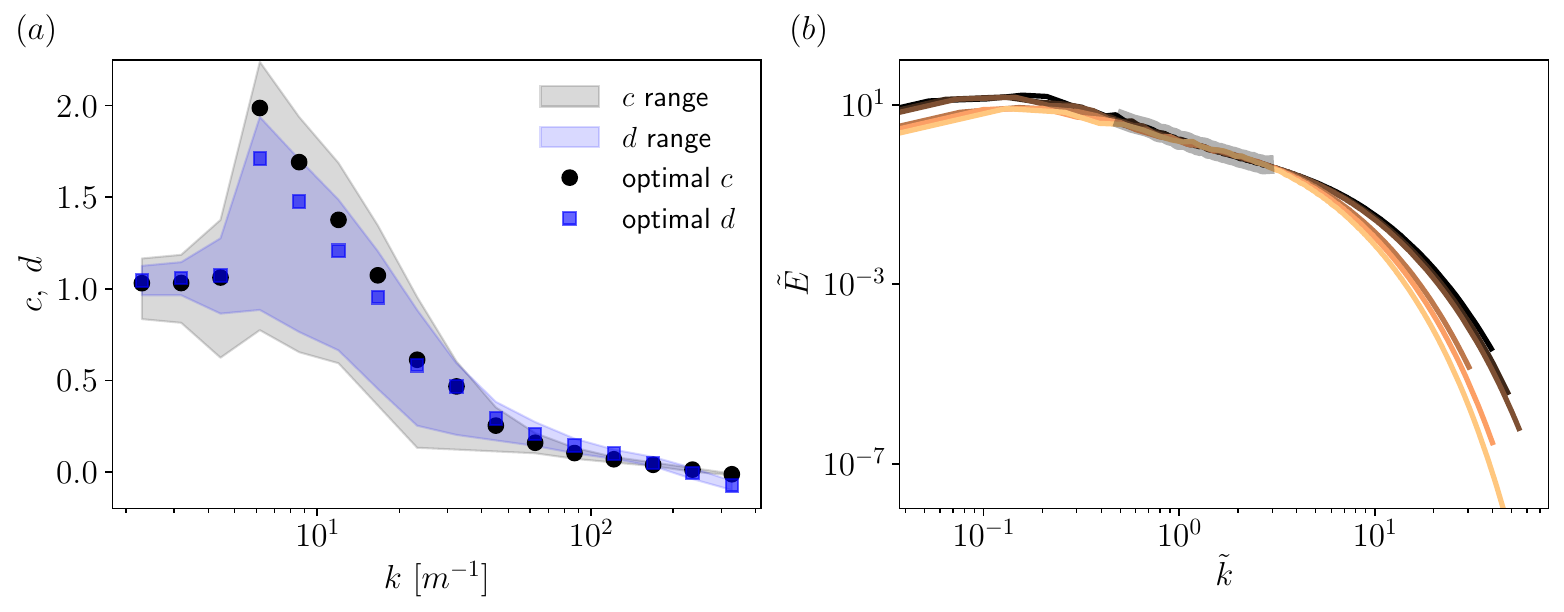}}
  \caption{$(a)$ Variation of the exponents $c$ (black circles) and $d$ (blue squares), as defined in \eqref{eq-dt-intermediate}, with the location of a fixed-length sliding window. The shaded areas highlight the admissible range of $c$ and $d$ values where $J<1.5J_m$ for each window. The window length is 0.71 decades. $(b)$ Rescaled energy spectra $\tilde{E}(\tilde{\kappa})$ using the expression identified for window 8, for which $J_m=0.0061$, obtained at $(c,d)=(0.613,0.583)$. The shaded region shows the collapsed range of the window.}
  \label{fig-dt_ab_y_4}
\end{figure}

Figure \ref{fig-dt_ab_y_4}$(a)$ shows the variation of the optimal exponents $c$ (black circles) and $d$ (blue squares) with window centre location. Their close agreement suggests self-similarity of expression \eqref{eq-dt-intermediate} for each window. As the window shifts from the inner to the outer region, the exponents first increase monotonically from 0, pass through a region where they exceed $[0,1]$, and ultimately return to 1. Figure \ref{fig-dt_ab_y_4}$(b)$ shows the rescaled energy spectra using the expression identified for window 8, located within the intermediate region. A good collapse is observed across the entire matched range. 
The sampling method is then applied to explore the ranges of $c$ and $d$ that can well collapse the data in each window. 
The shaded areas in figure \ref{fig-dt_ab_y_4}$(a)$ show the admissible ranges of $(c, d)$ satisfying $J<1.5J_m$ for each window, obtained using sampling methods. These ranges are substantially wider in the intermediate region than in the inner and outer regions, consistent with the behaviour observed in turbulent channel flow (figure \ref{fig-ch_ab_y_4}$(a)$). The main difference is that the leftmost window has moderately small, rather than extremely small, admissible ranges compared with turbulent channel flow, which can be attributed to the reasons discussed earlier in \S\ref{sec-dt-two-scale}.
Instead of varying the station set (i.e., changing the Reynolds number), we can also observe changes in the admissible ranges of $c$ and $d$ in the intermediate region by varying the window length. As the window length increases, the admissible ranges satisfying $J<1.5J_m$ narrow in the intermediate region, as shown in figures \ref{fig-dt_ab_y_3} and \ref{fig-dt_ab_y_2} in Appendix \ref{app-intermediate-fig}. This is because extending the window length with the intermediate-region extent fixed is equivalent to shrinking the intermediate region (i.e., reducing the Reynolds number) with the window length fixed. Both increase the ratio of the window length to the intermediate-region length, leaving less room for the scaling to vary --- hence restricting the range of admissible scaling behaviours.

Following an analogous derivation to the one presented in \S\ref{sec-ch-intermediate}, our data-driven method gives the following expression for the energy spectrum in the intermediate region:
\begin{gather}
  E \propto \epsilon^{2/3}\kappa^{-5/3}\left[\left(\kappa L\right)^{0.0809}\left(\kappa\eta\right)^{0.0983}\right]^{5/3}. \label{eq-dt-intermediate-corr}
\end{gather}
The mathematical details of the derivation are provided in Appendix \ref{app-dt-intermediate}.
Using $\eta/L = Re_L^{-3/4}$, where $Re_L = uL/\nu$ is the Reynolds number based on the integral length scale, equation \eqref{eq-dt-intermediate-corr} can be written as
\begin{gather}
  E \propto \epsilon^{2/3}\kappa^{-5/3}\left[Re_L^{-0.0737}\left(\kappa L\right)^{0.1792}\right]^{5/3}. \label{eq-dt-intermediate-corr-ReL}
\end{gather}
Equation \eqref{eq-dt-intermediate-corr-ReL} recovers the $-5/3$ scaling at leading order, with an additional correction factor, $(\kappa^{0.1792})^{5/3}$, that yields a flatter slope. The dashed and dash-dotted lines in figure \ref{fig-dt_one_scale}$(a)$ show the relations $E\propto\kappa^{-5/3}$ and $E\propto\kappa^{-0.8208\times5/3}$, respectively. Our result is in closer agreement with the slope of the energy spectrum in the intermediate region. This flatter slope is associated with the violation of the local equilibrium hypothesis of Kolmogorov's theory at relatively small wavenumbers \citep{steiros2022balanced}, for which $\Pi/\epsilon=\text{const.}<1$, with $\Pi$ being the interscale energy flux. On the basis of a balanced cascade assumption, in which the terms of the energy budget equation scale with each other, \citet{steiros2022balanced} proposed a correction to the $-5/3$ law that also captures the flatter slope in the intermediate range. In addition, the algorithm identifies a Reynolds-number correction, $Re_L^{-0.0737}$.

\subsection{Early-stage homogeneous decaying turbulence} \label{sec-decaying_turbulence_ic}

Over the past two decades, evidence from wind-tunnel experiments \citep{vassilicos2015dissipation} and numerical simulations \citep{goto2015energy,goto2016local,goto2016unsteady,meldi2018investigation,steiros2022turbulence,fang2023edqnm} has revealed a novel turbulence energy cascade that preserves the Kolmogorov $-5/3$ law over more than a decade of wavenumbers while following the scaling $C_\epsilon \propto Re_L^{-1}$. 
The physical mechanism underlying this scaling, however, remains unclear.
In this section, we examine the self-similarity of this cascade, attempting to  shed light on the origin of the nonclassical dissipation scaling.
We use the same DNS data as \citet{goto2016unsteady}, focusing on the early stage of decay (i.e. small decay time $\hat{t}$), where the nonclassical dissipation scaling emerges.

\begin{figure}
\centerline{\includegraphics[width=\textwidth]{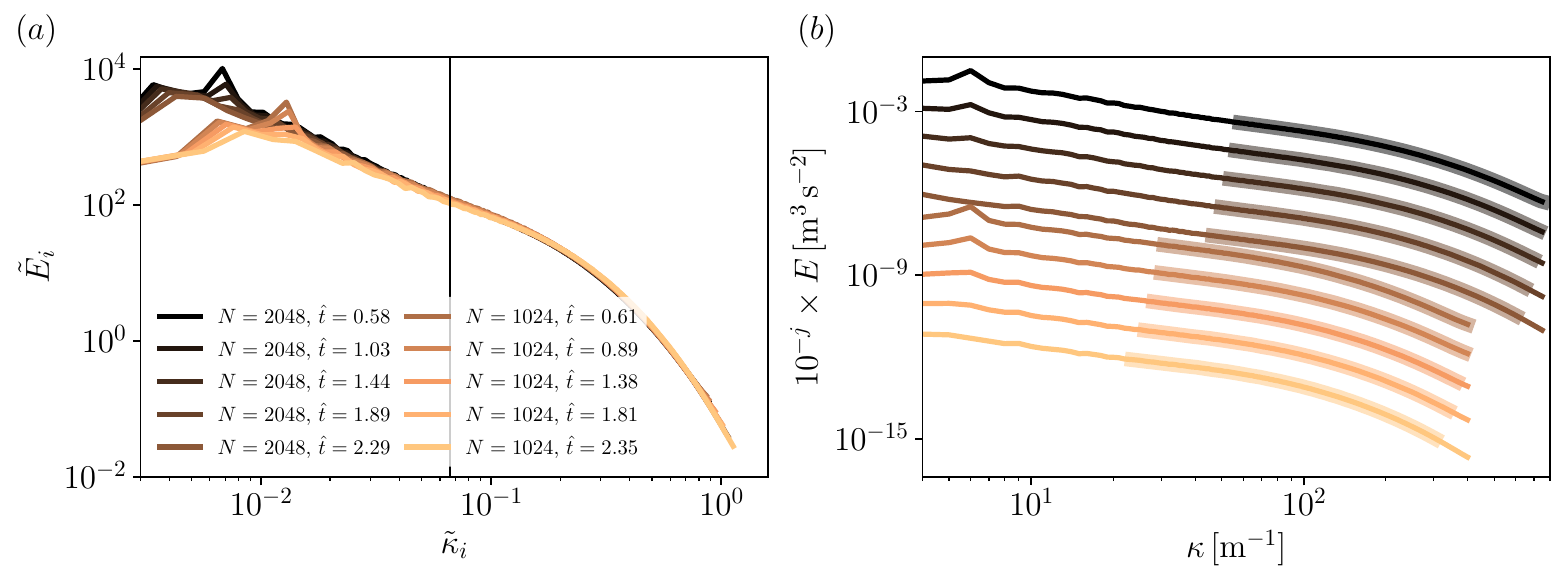}}
\caption{$(a)$ Rescaled energy spectra $\tilde{E}_i(\tilde{\kappa}_i)$ using the inner similarity expression \eqref{eq-dt-inner-ic}, identified for $\tilde{\kappa}_i>0.066$. The vertical line indicates $\tilde{\kappa}_i=0.066$. $(b)$ The corresponding collapsed range (shaded) of each profile for the inner similarity. The profiles are premultiplied by $10^{-j}$ for visual separation, where $j$ denotes the station index. Indices $j = 0$ to $4$ and $5$ to $9$ correspond to the cases $N=2048$ and $N=1024$, respectively, with each group ordered by increasing $\hat{t}$. The same line styles are used in figures \ref{fig-dt_outer_ic1}--\ref{fig-dt_outer_ic3}.}
\label{fig-dt_inner_ic}
\end{figure}

Ten stations with different decay times $\hat{t}$ are selected, five from each of the two simulation sizes $N=1024$ and $N=2048$.
The legend of figure \ref{fig-dt_inner_ic}$(a)$ lists the corresponding $\hat{t}$ and $N$ for each station.
With the profile $N = 2048$, $\hat{t} = 0.58$ chosen as the reference, the two-scale self-similarity framework identifies the same inner similarity as in the late stage (expression \eqref{eq-dt-inner}), scaled by $\epsilon$ and $\eta$, under the fixed-threshold criterion:
\begin{gather}
\text{Inner:}\quad \tilde{\kappa}_i = L^{0.016}\eta^{0.984}\kappa, \quad \tilde{E}_i = \epsilon^{-0.667}(L^{0.046}\eta^{0.954})^{-5/3}E. \label{eq-dt-inner-ic}
\end{gather}
This is in support of the classical Kolmogorov theory, which is shown to hold at sufficiently high wavenumbers even in strongly unsteady regimes.
The excellent collapse of the rescaled energy spectra and the corresponding collapsed range of each profile are shown in figure \ref{fig-dt_inner_ic}.

\begin{figure}
\centerline{\includegraphics[width=0.7\textwidth]{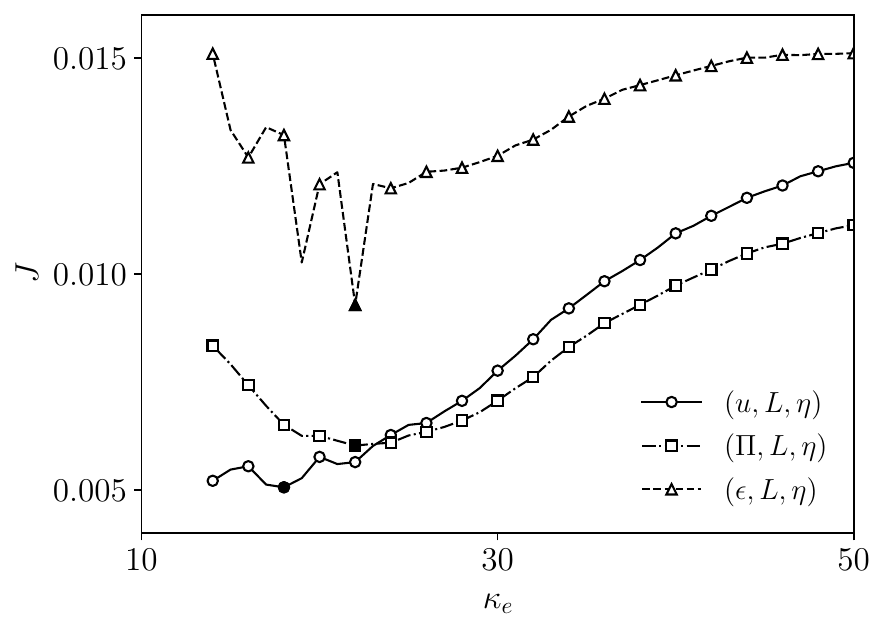}}
\caption{Variation of the objective function $J$ with the upper limit $\kappa_e$ of the search range $[8,\kappa_e]$. Solid, dash-dotted, and dashed lines correspond to outer similarity searches using scaling variables $(u, L, \eta)$, $(\Pi, L, \eta)$, and $(\epsilon, L, \eta)$, respectively. The global minimum of $J$ is indicated by the filled symbol.}
\label{fig-dt_loss_ic}
\end{figure}

As shown below, the outer-similarity range in the early stage of decay is substantially narrower than in the late stage, leaving fewer than 15 data points in the corresponding search range; identification of the outer similarity is therefore particularly challenging. We therefore restrict the transformation functions to power-law monomials of candidate variables informed by prior knowledge of the problem, namely the characteristic velocity $u$, the interscale energy flux $\Pi$, the dissipation rate $\epsilon$, the integral length scale $L$, and the Kolmogorov length scale $\eta$. Unlike the other quantities, which depend only on time $t$, $\Pi$ depends on both $\kappa$ and $t$. The global minimum criterion is used for identification as in the late-stage analysis (\S\ref{sec-dt-two-scale}).
Applying the algorithm to different libraries of candidate variables, we find that two expressions --- one scaled by $u$, $L$ and $\eta$, and the other by $\Pi$, $L$ and $\eta$ --- yield lower objective-function values than the expression scaled by $\epsilon$, $L$ and $\eta$, suggesting the breakdown of the $\epsilon$-based equilibrium normalisation in the early stage of decay.
Figure \ref{fig-dt_loss_ic} shows the variation of the objective function $J$ with the upper limit $\kappa_e$ of the search range $[8,\kappa_e]$ for these three cases, where the lower limit is set to 8 to exclude the influence of the initial forcing, which manifests as a spike in the energy spectrum.

\begin{figure}
\centerline{\includegraphics[width=\textwidth]{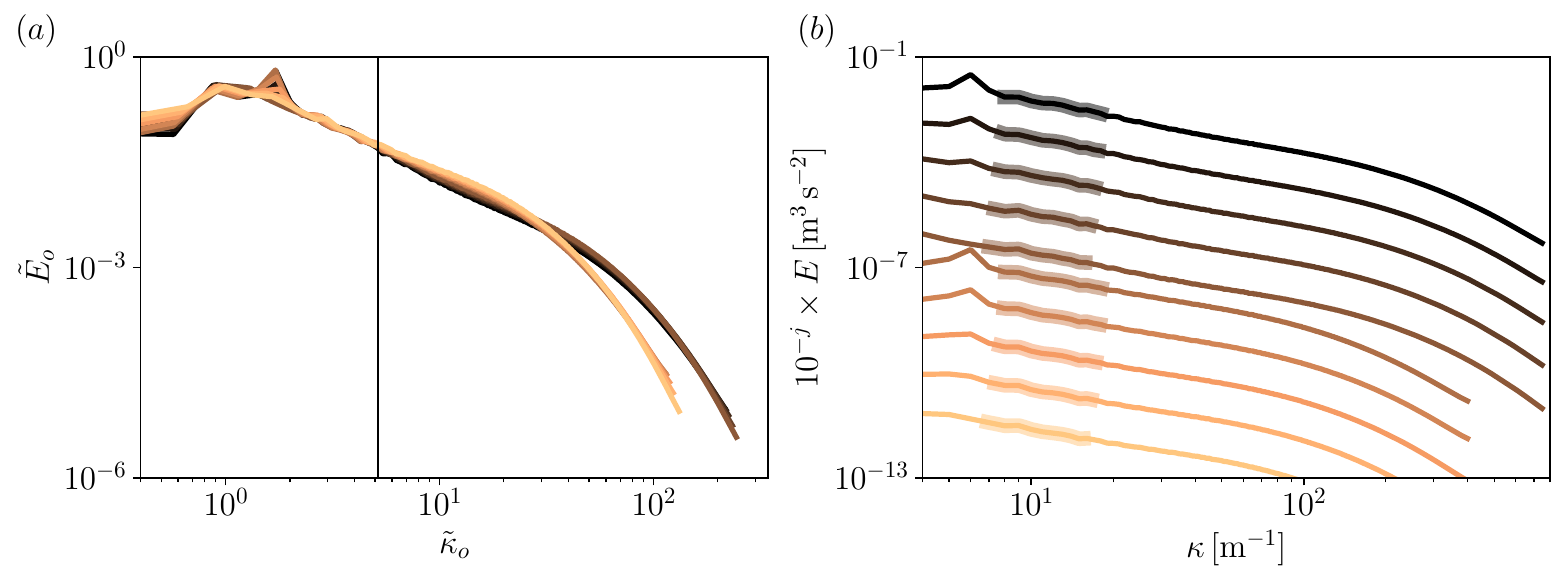}}
\caption{$(a)$ Rescaled energy spectra $\tilde{E}_o(\tilde{\kappa}_o)$ using the outer similarity expression \eqref{eq-dt-outer-ic1}, identified for $\tilde{\kappa}_o<5.147$. The vertical line indicates $\tilde{\kappa}_o = 5.147$. $(b)$ The corresponding collapsed range (shaded) of each profile for the outer similarity. The profiles are premultiplied by $10^{-j}$ for visual separation, where the station index $j$ is defined in the caption of figure \ref{fig-dt_inner_ic}.}
\label{fig-dt_outer_ic1}
\end{figure}

\begin{figure}
\centerline{\includegraphics[width=\textwidth]{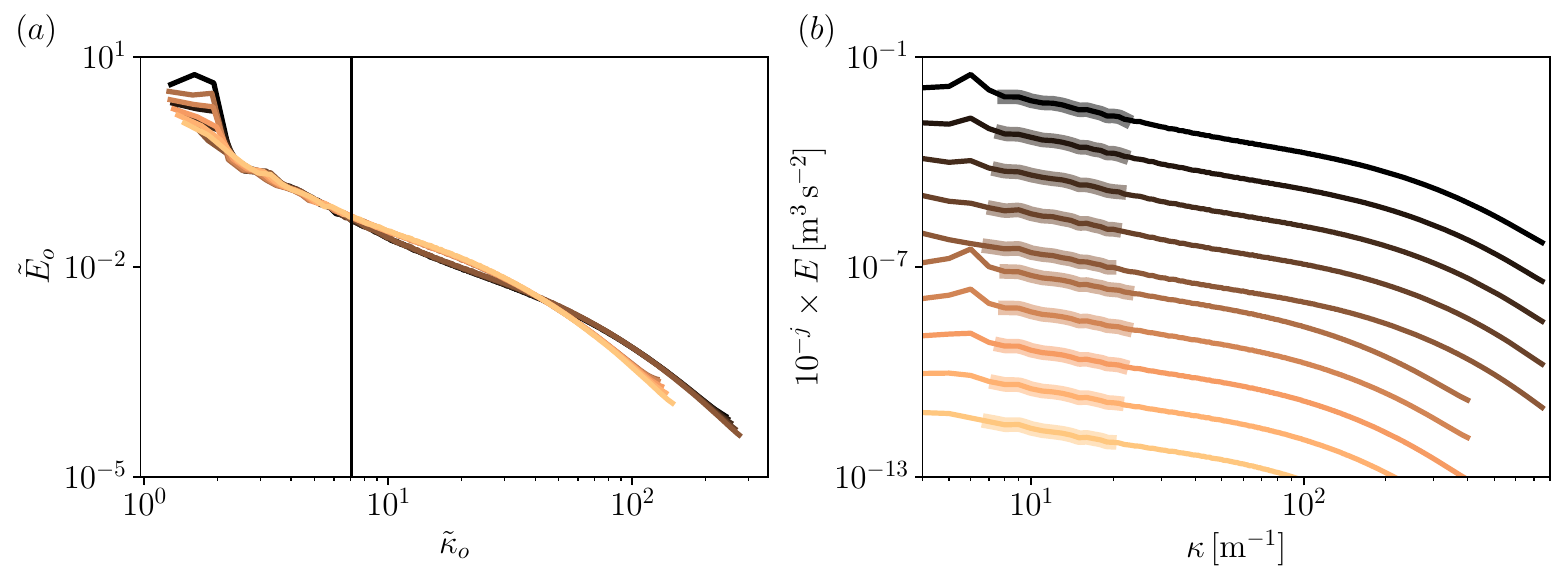}}
\caption{$(a)$ Rescaled energy spectra $\tilde{E}_o(\tilde{\kappa}_o)$ using the outer similarity expression \eqref{eq-dt-outer-ic2}, identified for $\tilde{\kappa}_o<7.077$. The vertical line indicates $\tilde{\kappa}_o = 7.077$. $(b)$ The corresponding collapsed range (shaded) of each profile for the outer similarity. The profiles are premultiplied by $10^{-j}$ for visual separation, where the station index $j$ is defined in the caption of figure \ref{fig-dt_inner_ic}.}
\label{fig-dt_outer_ic2}
\end{figure}

For the search using $u$, $L$ and $\eta$, the objective function $J$ decreases monotonically for $\kappa_e<50$ before weakly fluctuating at smaller $\kappa_e$, reaching a global minimum of 0.005 at $\kappa_e=18$ (filled circle). For the search using $\Pi$, $L$ and $\eta$, the value of $J$ first decreases monotonically to a global minimum of 0.006 at $\kappa_e=22$ (filled square), then increases monotonically as $\kappa_e$ is further reduced.
The similarity expressions corresponding to the two global minima are
\begin{align}
\smash{\raisebox{-0.6\baselineskip}{Outer:}}\quad
&\tilde{\kappa}_{o} = L^{0.970}\eta^{0.030}\kappa, \quad
\tilde{E}_{o} = u^{-2}(L^{1.008}\eta^{-0.008})^{-1}E, \label{eq-dt-outer-ic1}\\[-0.1ex]
&\tilde{\kappa}_{o} = L^{0.991}\eta^{0.009}\kappa, \quad
\tilde{E}_{o} = \Pi^{-2/3}(L^{1.027}\eta^{-0.027})^{-5/3}E. \label{eq-dt-outer-ic2}
\end{align}
Figures \ref{fig-dt_outer_ic1} and \ref{fig-dt_outer_ic2} show good collapse of the rescaled energy spectra using these two expressions, together with the corresponding collapsed range of each profile.
Expression \eqref{eq-dt-outer-ic1} was previously proposed through empirical collapse of energy spectra in homogeneous decaying turbulence generated by a fractal grid \citep{seoud2007dissipation,mazellier2010turbulence,george2009exponential}, where the dissipation scaling $C_\epsilon \propto Re_L^{-1}$ was first observed. In contrast to the present case, the turbulence in fractal-generated flows was proposed to be characterised by \eqref{eq-dt-outer-ic1} across the entire wavenumber range.
It is noteworthy that the objective functions of the two expressions exhibit a relatively similar variation with $\kappa_e$ (see figure \ref{fig-dt_loss_ic}). This may be explained by the fact that the interscale energy flux, $\Pi$, is closely proportional to $u^3/L$  at relatively low non-dimensional wavenumbers $\kappa L$ \citep{steiros2022turbulence}.

\begin{figure}
\centerline{\includegraphics[width=\textwidth]{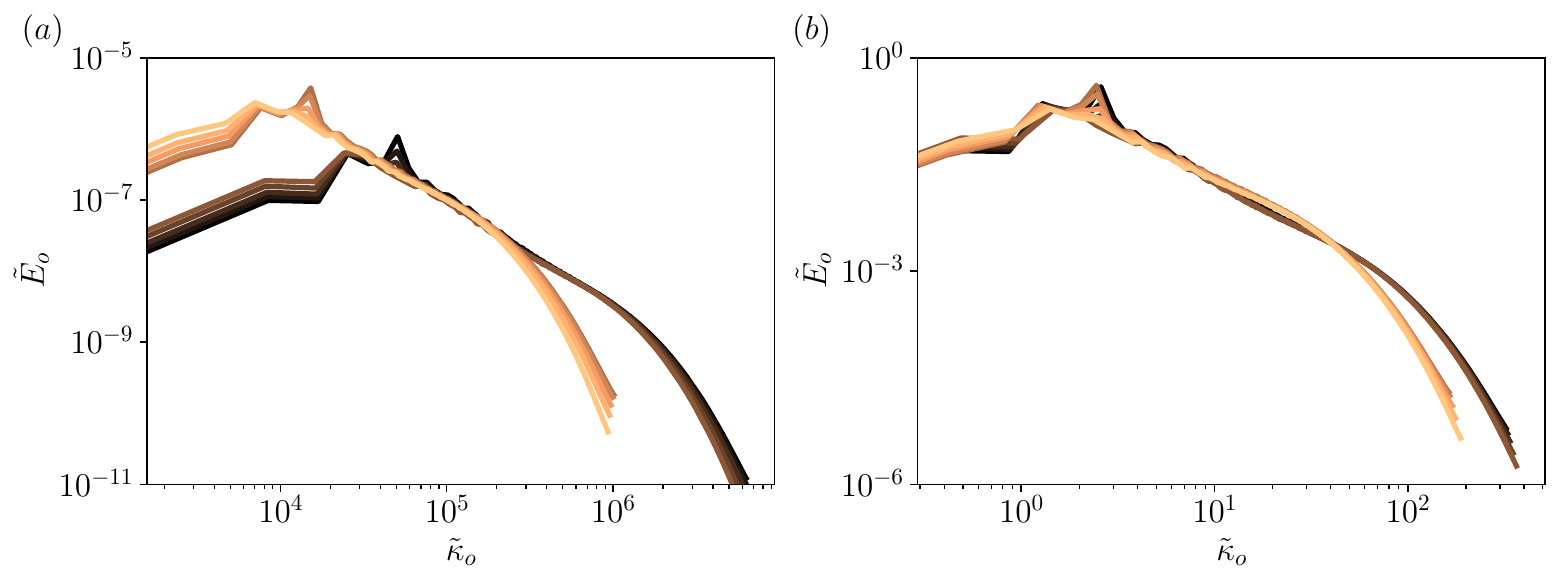}}
\caption{Rescaled energy spectra using outer similarity expressions scaled by $\epsilon$, $L$ and $\eta$, identified from two adjacent search ranges. $(a)$ Search range $[8,22]$: $\tilde{\kappa}_{o} = L^{2.747}\eta^{-1.747}\kappa$, $\tilde{E}_{o} = \epsilon^{-2/3}(L^{2.458}\eta^{-1.458})^{-5/3}E$. $(b)$ Search range $[8,23]$: $\tilde{\kappa}_{o} = L^{1.041}\eta^{-0.041}\kappa$, $\tilde{E}_{o} = \epsilon^{-2/3}(L^{1.098}\eta^{-0.098})^{-5/3}E$.}
\label{fig-dt_outer_ic3}
\end{figure}

For the expression scaled by $\epsilon$, $L$ and $\eta$, the objective function $J$ reaches a minimum of 0.009 at $\kappa_e = 22$ (filled triangle in figure \ref{fig-dt_loss_ic}), considerably larger than the minima obtained in the other two cases, and oscillates significantly for $\kappa_e<24$.
Figure \ref{fig-dt_outer_ic3} shows the rescaled energy spectra obtained from two adjacent search ranges, $[8,22]$ and $[8,23]$. Increasing $\kappa_e$ from 22 to 23 yields a markedly different outer similarity: in figure \ref{fig-dt_outer_ic3}$(a)$, the outer range of the energy spectra of the $N=2048$ case incorrectly matches the inertial range of the $N=1024$ case, producing a lower objective function value than the correct match shown in figure \ref{fig-dt_outer_ic3}$(b)$. This indicates that the outer similarity scaled by $\epsilon$, $L$ and $\eta$ is not robust to changes in $\kappa_e$, reinforcing the conclusion that a different scaling governs the outer similarity. 
This is further supported by the significant shrinkage of the outer-similarity range from the late stage ($\tilde{\kappa}<33.24$) to the early stage ($\tilde{\kappa}<5.147$ and $\tilde{\kappa}<7.077$). This shrinkage occurs despite the higher Reynolds number, which would otherwise be expected to broaden the range.

The breakdown of $\epsilon$-based outer scaling is also manifested in the self-similarity of the interscale energy flux $\Pi$. In spectral space, homogeneous decaying turbulence is governed by:
\begin{gather}
\frac{\partial K^>(\kappa,t)}{\partial t} = \Pi(\kappa,t) - \epsilon^>(\kappa,t), \label{eq-KHM}
\end{gather}
where $K^>(\kappa,t) = \int_\kappa^\infty Ed\kappa'$ and $\epsilon^>(\kappa,t) = 2\nu\int_\kappa^\infty \kappa'^2Ed\kappa'$ are the high-pass filtered kinetic energy and dissipation rate, respectively.
For equilibrium turbulence, $\partial K^>/\partial t \approx 0$, so $\Pi(\kappa)\approx\epsilon^>(\kappa)$. For scales that are large compared to the viscosity dominated scales ($\kappa\eta\to0$), viscous effects are negligible, $\epsilon^>(\kappa)\approx\epsilon$, and hence $\Pi$ is approximately equal to $\epsilon$. 
However, this result cannot be expected to hold at sufficiently large scales in the cascade, where the unsteady term $\partial K^>/\partial t$ becomes non-negligible.
Nevertheless, for the late stage of homogeneous decaying turbulence, \citet{goto2016unsteady} showed that in the limit of large eddies (i.e., $\kappa L \approx 1$), the unsteady term can be taken as proportional to the cascade dissipation. \citet{steiros2022balanced} generalised this result to a wide range of large-scale wavenumbers and derived the self-similarity expression $\Pi(\kappa,t) = \epsilon(t)g(\kappa L)$, where $g$ is a universal function. In the early stage of decay, however, this self-similar expression cannot be assumed to hold a priori, because the balanced nonequilibrium state of the large cascade eddies may be violated by the proximity to the initial conditions of the decay.

\begin{figure}
\centerline{\includegraphics[width=\textwidth]{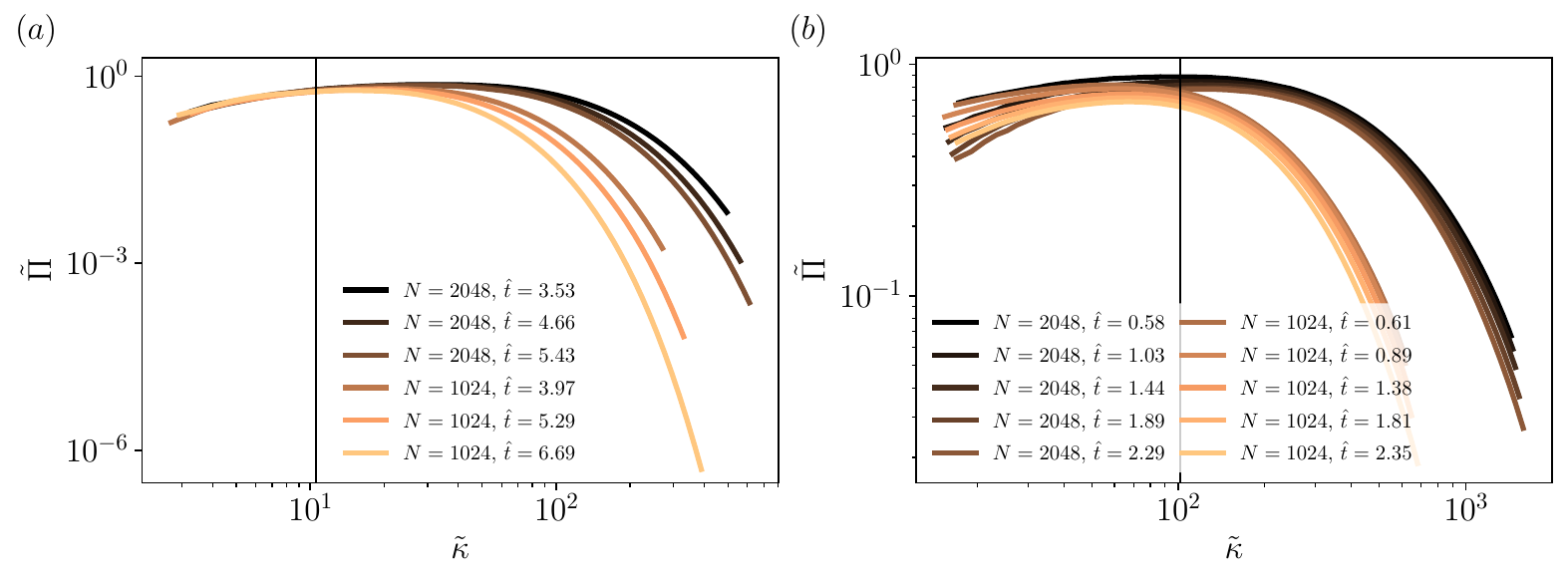}}
\caption{$(a)$ Rescaled interscale energy flux $\tilde{\Pi}(\tilde{\kappa})$ for the late stage of decay, using expression \eqref{eq-dt-Pi}, identified for $\tilde{\kappa}<10.573$. The vertical line marks $\tilde{\kappa}=10.573$. $(b)$ Rescaled interscale energy flux $\tilde{\Pi}(\tilde{\kappa})$ for the early stage of decay, using expression \eqref{eq-dt-Pi-ic}, identified for $\tilde{\kappa}<101.72$. The vertical line indicates $\tilde{\kappa}=101.72$.}
\label{fig-dt_Pi}
\end{figure}

In agreement with the above discussion, the algorithm successfully identifies a self-similarity for the interscale energy flux $\Pi$ in the late stage of decay, but not in the early one.
The late-stage outer similarity expression is found to be
\begin{gather}
\text{Late stage:}\quad \tilde{\kappa} = L^{1.048}\eta^{-0.048}\kappa, \quad \tilde{\Pi} = \epsilon^{-1}\Pi, \label{eq-dt-Pi}
\end{gather}
This expression is obtained from the range $[4,\kappa_e]$ at the reference station ($N = 2048$, $\hat{t} = 3.53$), with $\kappa_e=16$, where the objective function reaches its global minimum, $J=0.002$. The data for $\kappa<4$ are discarded because $\Pi$ is very close to zero in this range.
For the early stage, the expression is obtained from the range $[8,\kappa_e]$ at the reference station ($N = 2048$, $\hat{t} = 0.58$), with $\kappa_e = 53$, and is given by
\begin{gather}
\text{Early stage:}\quad \tilde{\kappa} = L^{1.299}\eta^{-0.299}\kappa, \quad \tilde{\Pi} = \epsilon^{-1}\Pi. \label{eq-dt-Pi-ic}
\end{gather}
Figure \ref{fig-dt_Pi} shows the rescaled interscale energy flux $\tilde{\Pi}$ for the late and early stages of decay using expressions \eqref{eq-dt-Pi} and \eqref{eq-dt-Pi-ic}, respectively. The late stage yields an excellent collapse of $\tilde{\Pi}$, whereas the early-stage profiles do not collapse. This is consistent with \citet{steiros2022turbulence}, where a new, non self-similar expression for $\Pi/\epsilon$ was derived for the early stage of decay.

The above results reinforce the view that, in the early stage of decay, the departure from the classical dissipation scaling $C_\epsilon = \text{const.}$ stems from the breakdown (or incomplete development) of a balanced nonequilibrium evolution of the large scales of the cascade, in which $\Pi\propto\epsilon$. This breakdown is reflected in both the loss of the $\epsilon$-based outer similarity and the pronounced shrinkage of the outer-similarity range.
It is noteworthy that, since the inner similarity region of the early decay is governed by $\epsilon$ and $\eta$ (same as the late decay), the nonclassical dissipation scaling is unlikely to be attributed to anomalies in the small scales of the cascade.

\section{Summary}\label{sec-sum}
A data-driven methodology has been developed for the systematic identification of multiscale self-similarity, building on the single-scale framework proposed by \citet{bempedelis2025extracting}.
Rather than analysing the entire range of data across all stations, the multiscale framework selects a specific range at a reference station and formulates an optimisation problem to identify the corresponding ranges at other stations that best align with it.
If self-similarity is present, the analytic form of the similarity transformation can then be recovered using symbolic regression.
On this basis, a workflow is developed to systematically identify possible multiscale self-similarity through an iterative search process.
The proposed methodology differs from previous data-driven approaches to self-similarity identification, which are restricted to a single scale, and broadens the scope of analysis to multiscale problems commonly encountered in fluid mechanics and many other fields.

The methodology is implemented as an algorithm and first applied to two canonical fluid-mechanical problems: turbulent channel flow and late-stage homogeneous decaying turbulence characterised by the classical dissipation law. 
In turbulent channel flow, the algorithm identifies two-scale self-similarity in the mean velocity gradient. The inner and outer regions are characterised by the viscous length scale $\delta_\nu$ and the channel half-height $\delta$, respectively, with the friction velocity $u_\tau$ providing the velocity scale in both regions.
In late-stage homogeneous decaying turbulence, the algorithm similarly identifies two-scale self-similarity in the energy spectrum, with the inner region characterised by the Kolmogorov length scale $\eta$ and the outer region by the integral length scale $L$, while the dissipation rate $\epsilon$ characterises both regions.
In the intermediate regions, where the inner and outer similarities overlap, the algorithm identifies similarity expressions that are independent of both the inner and outer length scales. These expressions lead to the corresponding scaling laws: the logarithmic law for the mean velocity profile in turbulent channel flow and the $-5/3$ law for the energy spectrum in late-stage homogeneous decaying turbulence, providing a new data-driven route to recovering classical scaling laws.
The same data-driven procedure also reveals departures from the leading-order laws, uncovering corrections to both cases: one linked to the power-law scaling proposed by \citet{barenblatt1993scalinga,barenblatt1993scalingb}, and the other yielding an improved approximation of the energy spectrum, further demonstrating that the framework can identify not only leading-order scaling laws but also higher-order corrections.

The algorithm is then applied to early-stage homogeneous decaying turbulence, where the nonclassical dissipation law is observed. In this case, it identifies the same inner similarity as in late-stage decay, while discovering two distinct outer similarity expressions: one characterised by the velocity scale $u$ and $L$, and the other by the interscale energy flux $\Pi$ and $L$. Both achieve better and more robust collapse than the expression characterised by $\epsilon$ and $L$ used in the late stage.  
No matter the choice of similarity variables, it was observed that the range of scales that exhibited outer similarity in the early decay was considerably reduced compared to the late, balanced nonequilibrium decay.

Although the multiscale self-similarity framework proposed in this paper is demonstrated on problems involving two governing parameters, it can be readily extended to higher-dimensional settings and more complex similarity transformations, including rotation, reflection, scaling and shearing, by representing similarity variables in the form of $\boldsymbol{\xi} =\boldsymbol{A}(t)\boldsymbol{s} + \boldsymbol{b}(t)$ and $\tilde{q} = \gamma(t) q(\boldsymbol{s},t) + \delta(t)$,
where $\boldsymbol{s}$ and $t$ are governing parameters, $\boldsymbol{\xi}$ denotes the similarity variables, $\boldsymbol{A}$ is a linear transformation matrix, and $\boldsymbol{b}$ is a translation vector. Bold symbols denote vectors and matrices. More broadly, the proposed framework is applicable to a wide range of problems in physics, engineering and beyond.


\begin{bmhead}[Acknowledgements.] 
The authors are indebted to Prof. Susumu Goto and Prof. John Christos Vassilicos for generously providing the DNS data. 
K.Z. acknowledges Prof. Xuesong Wu for the inspiring discussions.
\end{bmhead}

\begin{bmhead}[Funding.]
K.Z. and K.S. acknowledge support from the ERC Starting Grant ONSET No. 10116332
\end{bmhead}

\begin{bmhead}[Declaration of interests.]
The authors report no conflict of interest.
\end{bmhead}

\begin{bmhead}[Data availability statement.]
The code and input data required to reproduce the work reported in Section \ref{sec-turbulent_channel} will be made available on GitHub after the peer-review process is complete.
\end{bmhead}


\appendix
\section{Sensitivity analysis of inner and outer similarity in turbulent channel flow} \label{app-ch-sensitivity}

This section examines the sensitivity of the turbulent-channel-flow similarity expressions to the choices of station set, reference station, and grid distribution.
Tables \ref{table-ch-stations}--\ref{table-ch-grid} summarise the inner and outer similarity expressions obtained by applying the two-scale analysis under these different choices. The station index $j$ is defined in the caption of figure \ref{fig-ch_one_scale}. The resulting expressions are very close to one another, indicating that the algorithm is robust to these choices.

\begin{table}
  \centering
  \small
  \setlength{\tabcolsep}{5pt}
  \renewcommand{\arraystretch}{1.9}
  \begin{tabular}{>{\centering\arraybackslash}m{0.13\textwidth}
                  *{4}{>{\centering\arraybackslash}m{0.18\textwidth}}}
    \toprule
    \multirow{2}{*}{\shortstack{Stations\\(index $j$)}}  & \multicolumn{2}{c}{Inner similarity} & \multicolumn{2}{c}{Outer similarity}  \\
    \cmidrule(lr){2-3}\cmidrule(lr){4-5}
    & $\tilde{y}_i/y$                      & $\left.\widetilde{\frac{\mathrm{d}\bar{U}}{\mathrm{d}y}}\right|_i/\frac{\mathrm{d}\bar{U}}{\mathrm{d}y}$ & $\tilde{y}_o/y$                      & $\left.\widetilde{\frac{\mathrm{d}\bar{U}}{\mathrm{d}y}}\right|_o/\frac{\mathrm{d}\bar{U}}{\mathrm{d}y}$ \\
    \midrule
    \shortstack{$[0,1,2,3,4,$\\$5,6,7,8,9]$}   & $\delta^{0.008}\delta_\nu^{-1.008}$ & $u_\tau^{-1}\delta^{0.001}\delta_\nu^{0.999}$   & $\delta^{-1.000}\delta_\nu^{-0.000}$ & $u_\tau^{-1}\delta^{0.962}\delta_\nu^{0.038}$                                                            \\
    $[0,2,4,5,7,9]$   & $\delta^{0.018}\delta_\nu^{-1.018}$ & $u_\tau^{-1}\delta^{0.000}\delta_\nu^{1.000}$   & $\delta^{-1.000}\delta_\nu^{-0.000}$ & $u_\tau^{-1}\delta^{0.962}\delta_\nu^{0.038}$                                                            \\
    $[0,1,2,5,6,7]$   &   $\delta^{0.014}\delta_\nu^{-1.014}$ & $u_\tau^{-1}\delta^{0.001}\delta_\nu^{0.999}$   & $\delta^{-0.999}\delta_\nu^{-0.001}$  & $u_\tau^{-1}\delta^{1.030}\delta_\nu^{-0.030}$                                                           \\
    $[0,1,2,3,4]$   &   $\delta^{0.016}\delta_\nu^{-1.016}$  &   $u_\tau^{-1}\delta^{0.000}\delta_\nu^{1.000}$   &  $\delta^{-1.000}\delta_\nu^{0.000}$ & $u_\tau^{-1}\delta^{0.960}\delta_\nu^{0.040}$                                                            \\
    $[5,6,7,8,9]$   &  $\delta^{0.017}\delta_\nu^{-1.017}$   &   $u_\tau^{-1}\delta^{0.001}\delta_\nu^{0.999}$   &   $\delta^{-1.000}\delta_\nu^{0.000}$  &   $u_\tau^{-1}\delta^{0.966}\delta_\nu^{0.034}$                                                            \\
    $[0,2,4]$  &  $\delta^{0.018}\delta_\nu^{-1.018}$  &   $u_\tau^{-1}\delta^{0.000}\delta_\nu^{1.000}$   &   $\delta^{-1.000}\delta_\nu^{0.000}$ & $u_\tau^{-1}\delta^{0.961}\delta_\nu^{0.039}$                                                            \\
    \bottomrule
  \end{tabular}
  \caption{Inner and outer similarity expressions identified by choosing different station sets. The station with the lowest $Re_\tau$ is chosen as the reference station in each set.}
  \label{table-ch-stations}
\end{table}

\begin{table}
  \centering
  \small
  \setlength{\tabcolsep}{5pt}
  \renewcommand{\arraystretch}{1.9}
  \begin{tabular}{>{\centering\arraybackslash}m{0.13\textwidth}
                  *{4}{>{\centering\arraybackslash}m{0.18\textwidth}}}
    \toprule
    \multirow{2}{*}{\shortstack{Reference station\\(index $j$)}} & \multicolumn{2}{c}{Inner similarity} & \multicolumn{2}{c}{Outer similarity}  \\
    \cmidrule(lr){2-3}\cmidrule(lr){4-5}
    & $\tilde{y}_i/y$      & $\left.\widetilde{\frac{\mathrm{d}\bar{U}}{\mathrm{d}y}}\right|_i/\frac{\mathrm{d}\bar{U}}{\mathrm{d}y}$ & $\tilde{y}_o/y$     & $\left.\widetilde{\frac{\mathrm{d}\bar{U}}{\mathrm{d}y}}\right|_o/\frac{\mathrm{d}\bar{U}}{\mathrm{d}y}$ \\
    \midrule
    $0$   & $\delta^{0.008}\delta_\nu^{-1.008}$ & $u_\tau^{-1}\delta^{0.001}\delta_\nu^{0.999}$   & $\delta^{-1.000}\delta_\nu^{-0.000}$ & $u_\tau^{-1}\delta^{0.962}\delta_\nu^{0.038}$                                                            \\
    $2$   & $\delta^{-0.018}\delta_\nu^{-0.982}$ & $u_\tau^{-1}\delta^{-0.001}\delta_\nu^{1.001}$   & $\delta^{-1.000}\delta_\nu^{0.000}$ & $u_\tau^{-1}\delta^{0.962}\delta_\nu^{0.038}$                                                            \\
    $4$   & $\delta^{0.003}\delta_\nu^{-1.003}$ & $u_\tau^{-1}\delta^{0.004}\delta_\nu^{0.996}$    & $\delta^{-1.000}\delta_\nu^{0.000}$ & $u_\tau^{-1}\delta^{0.962}\delta_\nu^{0.038}$                                                            \\
    $5$   & $\delta^{0.004}\delta_\nu^{-1.004}$ & $u_\tau^{-1}\delta^{0.004}\delta_\nu^{0.996}$    & $\delta^{-1.000}\delta_\nu^{0.000}$ & $u_\tau^{-1}\delta^{0.963}\delta_\nu^{0.037}$                                                            \\
    \bottomrule
  \end{tabular}
  \caption{Inner and outer similarity expressions identified by choosing different reference stations.}
  \label{table-ch-reference}
\end{table}

\begin{table}
  \centering
  \small
  \setlength{\tabcolsep}{5pt}
  \renewcommand{\arraystretch}{1.9}
  \begin{tabular}{>{\centering\arraybackslash}m{0.13\textwidth}
                  *{4}{>{\centering\arraybackslash}m{0.18\textwidth}}}
    \toprule
    \multirow{2}{*}{\shortstack{Grid distribution}} & \multicolumn{2}{c}{Inner similarity} & \multicolumn{2}{c}{Outer similarity}  \\
    \cmidrule(lr){2-3}\cmidrule(lr){4-5}
    & $\tilde{y}_i/y$      & $\left.\widetilde{\frac{\mathrm{d}\bar{U}}{\mathrm{d}y}}\right|_i/\frac{\mathrm{d}\bar{U}}{\mathrm{d}y}$ & $\tilde{y}_o/y$     & $\left.\widetilde{\frac{\mathrm{d}\bar{U}}{\mathrm{d}y}}\right|_o/\frac{\mathrm{d}\bar{U}}{\mathrm{d}y}$ \\
    \midrule
    linear   & $\delta^{0.021}\delta_\nu^{-1.021}$ & $u_\tau^{-1}\delta^{0.023}\delta_\nu^{0.977}$   & $\delta^{-1.000}\delta_\nu^{-0.000}$ & $u_\tau^{-1}\delta^{0.962}\delta_\nu^{0.038}$                                                            \\
    logarithmic   & $\delta^{0.008}\delta_\nu^{-1.008}$ & $u_\tau^{-1}\delta^{0.001}\delta_\nu^{0.999}$    & $\delta^{-1.000}\delta_\nu^{0.000}$ & $u_\tau^{-1}\delta^{0.963}\delta_\nu^{0.037}$                                                            \\                                                     \\
    \bottomrule
  \end{tabular}
  \caption{Inner and outer similarity expressions identified by choosing different grid distributions.}
  \label{table-ch-grid}
\end{table}



\section{Outer similarity characterised by $U_c-\bar{U}$ in turbulent channel flow} \label{app-ch-outer}

The velocity scale, $U_c-\bar{U}$, proposed by \citet{zagarola1997scaling, zagarola1998mean} for the outer similarity in turbulent pipe flow is examined for turbulent channel flow. The following expression is obtained when all ten stations are included in the analysis:
\begin{gather}
  \tilde{y}_o = \delta^{-1.000}\delta_\nu^{0.000}y, \quad  \left.\widetilde{\frac{\mathrm{d}\bar{U}}{\mathrm{d}y}}\right|_o = (U_c-\bar{U})^{-1}\delta^{0.963}\delta_\nu^{0.037}\frac{\mathrm{d}\bar{U}}{\mathrm{d}y}, \label{eq-ch-outer-zs1}
\end{gather}
The power-law exponents of the two length scales in \eqref{eq-ch-outer-zs1} are close to those in \eqref{eq-ch-outer-1}, where $u_\tau$ is used as the velocity scale. These results agree with \citet{pirozzoli2023outer}, who observed a similar level of collapse in turbulent channel flow when using $U_c-\bar{U}$ and $u_\tau$ as velocity scales.

\section{Effect of Reynolds number on the ranges of inner and outer similarity in turbulent channel flow} \label{app-ch-reynolds}

In this section, we examine how the inner and outer similarity ranges depend on the station with the lowest $Re_\tau$.
We compare the second case in table \ref{table-ch-stations}, where only the six stations with the highest $Re_\tau$ are included, with the case discussed in \S \ref{sec-ch-two-scale}, where all ten stations are included.

\begin{figure}
  \centerline{\includegraphics[width=\textwidth]{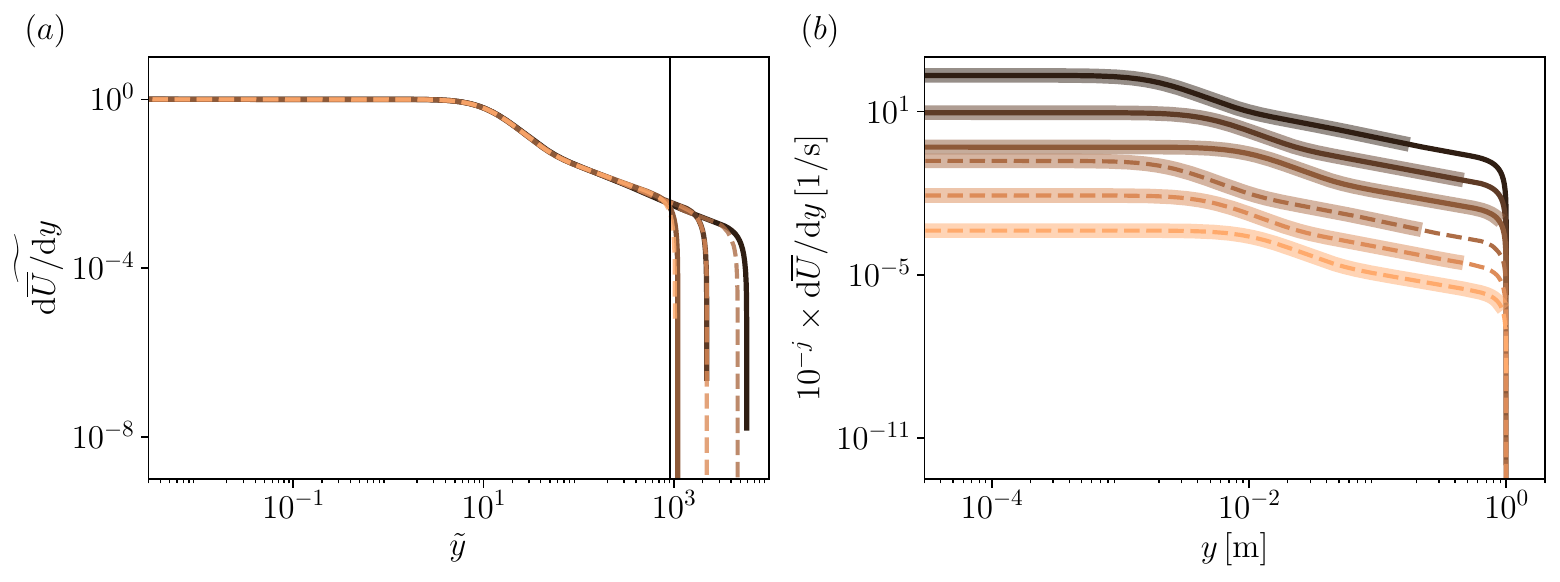}}
  \caption{$(a)$ Rescaled mean velocity gradient profiles using the inner similarity expression in table \ref{table-ch-stations}, identified for $\tilde{y} < 908.0$. The vertical line indicates $\tilde{y} = 908.0$. $(b)$ The corresponding collapsed ranges (shaded) of profiles for the inner similarity. The profiles are premultiplied by $10^{-j}$ for visual separation, where the station index $j$ is defined in the caption of figure \ref{fig-ch_one_scale}.}
  \label{fig-ch_inner_2}
\end{figure}

\begin{figure}
  \centerline{\includegraphics[width=\textwidth]{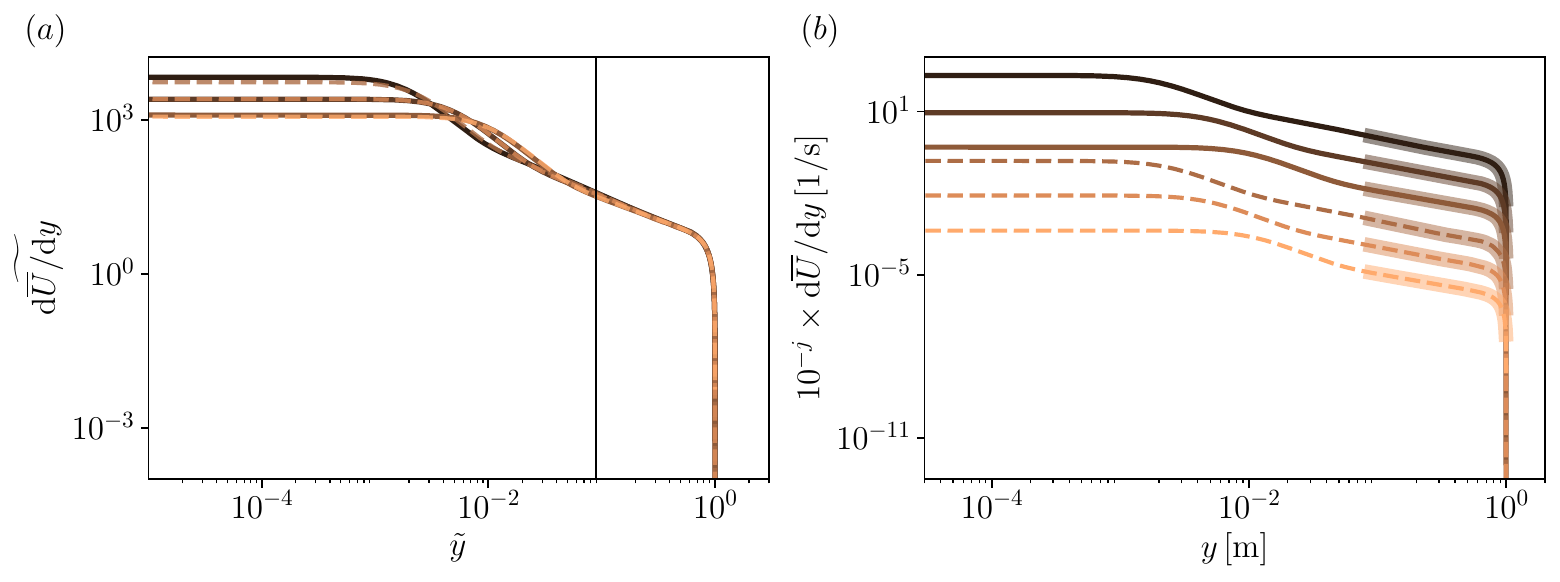}}
  \caption{$(a)$ Rescaled mean velocity gradient profiles using the outer similarity expression in table \ref{table-ch-stations}, identified for $\tilde{y} > 0.090$. The vertical line indicates $\tilde{y} = 0.090$. $(b)$ The corresponding collapsed ranges (shaded) of profiles for the outer similarity. The profiles are premultiplied by $10^{-j}$ for visual separation, where the station index $j$ is defined in the caption of figure \ref{fig-ch_one_scale}.}
  \label{fig-ch_outer_2}
\end{figure}

Figures \ref{fig-ch_inner_2}$(a)$ and \ref{fig-ch_outer_2}$(a)$ show successful collapse of the six stations in the rescaled coordinates using the corresponding inner and outer similarity expressions.
The inner and outer similarities are identified in the regions $\tilde{y}_i<908.0$ and $\tilde{y}_o>0.090$, respectively, as indicated by the vertical lines.
The corresponding collapsed regions for each station are displayed as shaded areas in figures \ref{fig-ch_inner_2}$(b)$ and \ref{fig-ch_outer_2}$(b)$ by rescaling $\tilde{y}_i$ and $\tilde{y}_o$ back to $y$. Compared with figures \ref{fig-ch_inner_1}$(b)$ and \ref{fig-ch_outer_1}$(b)$, both the inner and outer collapsed regions of the six stations are significantly extended because the low-$Re_\tau$ stations are excluded. The existence of an overlap is again examined at the lowest $Re_\tau$ station, $Re_\tau=934$, as in this case the limits of the two ranges are determined by this station.
The inner and outer regions in the $y$ scale correspond to $y<0.882$ and $y>0.090$, which overlap. The overlap region obtained for $Re_\tau = 934$ here is larger than that for $Re_\tau = 182$ obtained in \S\ref{sec-ch-two-scale}, as expected with increasing $Re_\tau$.

\section{Further analysis of the intermediate region}\label{app-intermediate-fig}

\begin{figure}
  \centerline{\includegraphics[width=\textwidth]{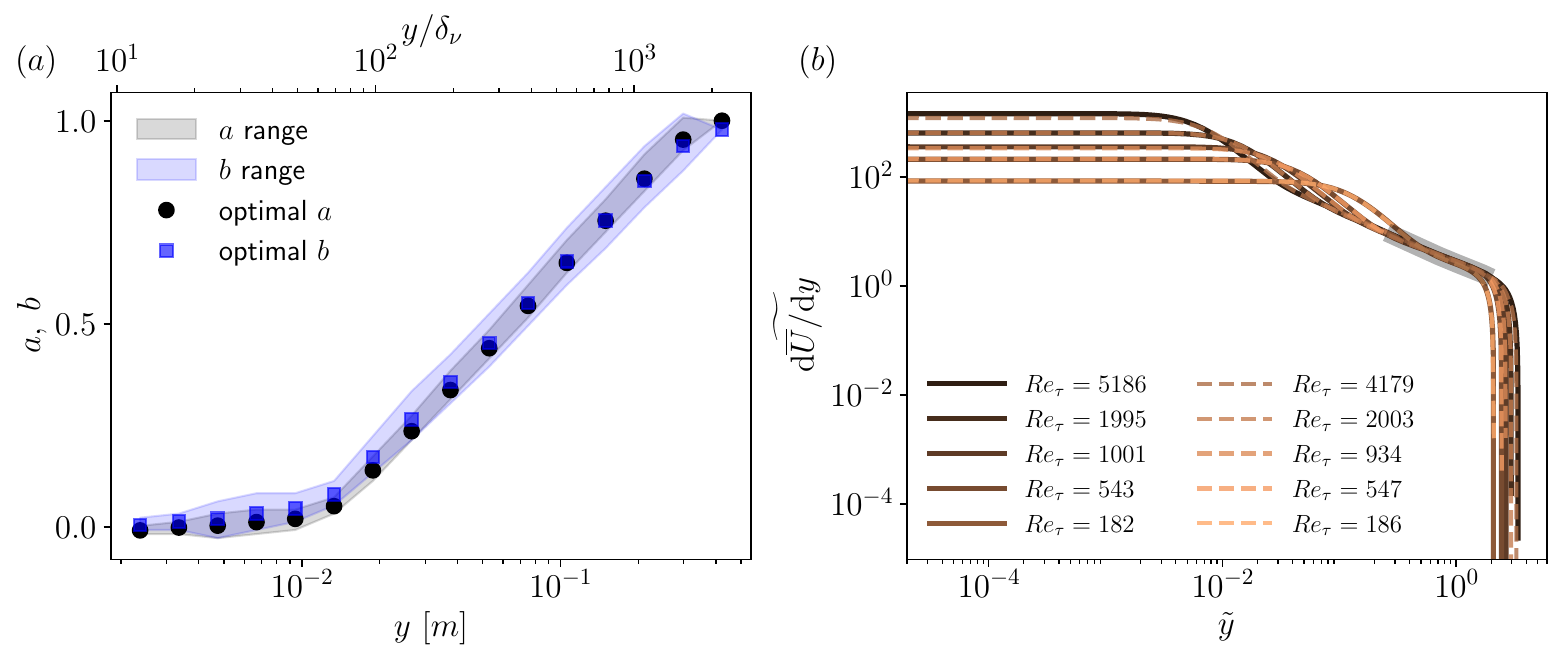}}
  \caption{$(a)$ Variation of the exponents $a$ (black circles) and $b$ (blue squares), as defined in \eqref{eq-ch-intermediate}, with the location of a fixed-length sliding window. The shaded areas highlight the admissible range of $a$ and $b$ values where $J<1.5J_m$ for each window. $(b)$ Rescaled mean velocity gradient profiles using the expression identified for window 14, which has the largest $J_m$ value of 0.0121 among all windows, obtained at $(a,b)=(0.857,0.851)$. The shaded region shows the collapsed range of the window. All ten profiles are included in the analysis.}
  \label{fig-ch_ab_y_10}
\end{figure}

\begin{figure}
  \centerline{\includegraphics[width=\textwidth]{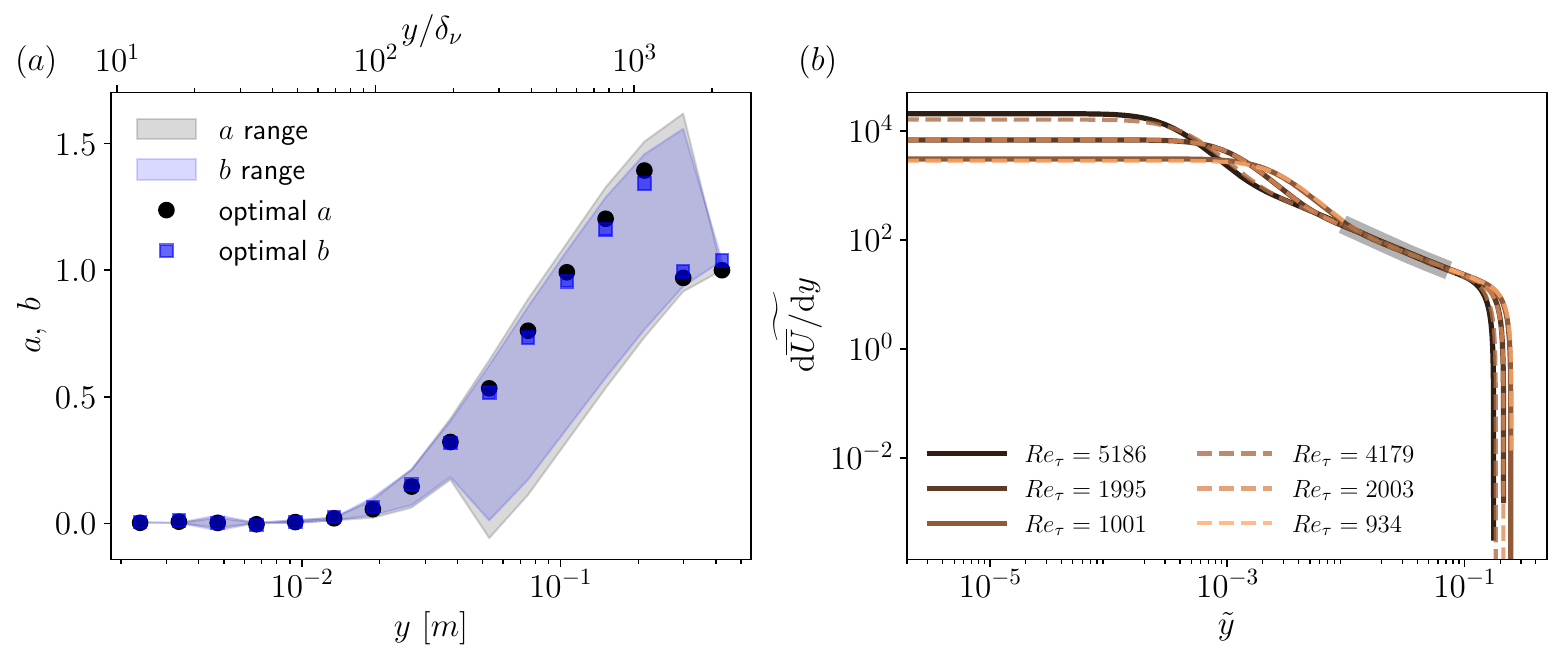}}
  \caption{$(a)$ Variation of the exponents $a$ (black circles) and $b$ (blue squares), as defined in \eqref{eq-ch-intermediate}, with the location of a fixed-length sliding window. The shaded areas highlight the admissible range of $a$ and $b$ values where $J<1.5J_m$ for each window. $(b)$ Rescaled mean velocity gradient profiles using the expression identified for window 13, which has the largest $J_m$ value of 0.0020 among all windows, obtained at $(a,b)=(1.203,1.161)$. The shaded region shows the collapsed range of the window. Six profiles are included in the analysis.}
  \label{fig-ch_ab_y_6}
\end{figure}

For the case of turbulent channel flow, figures \ref{fig-ch_ab_y_10} and \ref{fig-ch_ab_y_6} show the variation of the exponents $a$ and $b$, defined in \eqref{eq-ch-intermediate}, as a function of the sliding window centre location, for all ten stations and for the six highest-$Re_\tau$ stations, respectively. The shaded areas show the admissible ranges of $(a, b)$ satisfying $J<1.5J_m$ for each window.

In figure \ref{fig-ch_ab_y_10}$(a)$, both exponents increase monotonically from 0 to 1 with the window centre location, which differs from the behaviour in figure \ref{fig-ch_ab_y_4}$(a)$ where only four profiles with highest $Re_\tau$ are included. However, within the intermediate range, the objective function values are significantly larger than those in the inner and outer regions. Figure \ref{fig-ch_ab_y_10}$(b)$ highlights the collapsed range of window 14, which has the largest $J_m$ value of 0.0121 among all windows. A good collapse is observed in the middle of the range but deteriorates toward the two ends, primarily because the two lowest-$Re_\tau$ stations (182 and 186, which have the shortest overlap region) do not provide a sufficiently long range for collapse. The limited extent of the intermediate region further constrains the admissible range of $a$ and $b$. Figure \ref{fig-ch_ab_y_6} shows the results for the six highest-$Re_\tau$ stations. The monotonic variation of $a$ and $b$ is disrupted by excluding the low-$Re_\tau$ stations, and the admissible ranges are wider than in figure \ref{fig-ch_ab_y_10}$(a)$ but narrower than in figure \ref{fig-ch_ab_y_4}$(a)$.
These data-driven results suggest the existence of an intermediate region at sufficiently high Reynolds numbers, in which the profiles can be collapsed using arbitrary values of $a$ and $b$, provided that $a \approx b$. This suggests that no characteristic length scale is distinguished within this region, consistent with the scale-free nature of the logarithmic law governing the intermediate region.

\begin{figure}
  \centerline{\includegraphics[width=\textwidth]{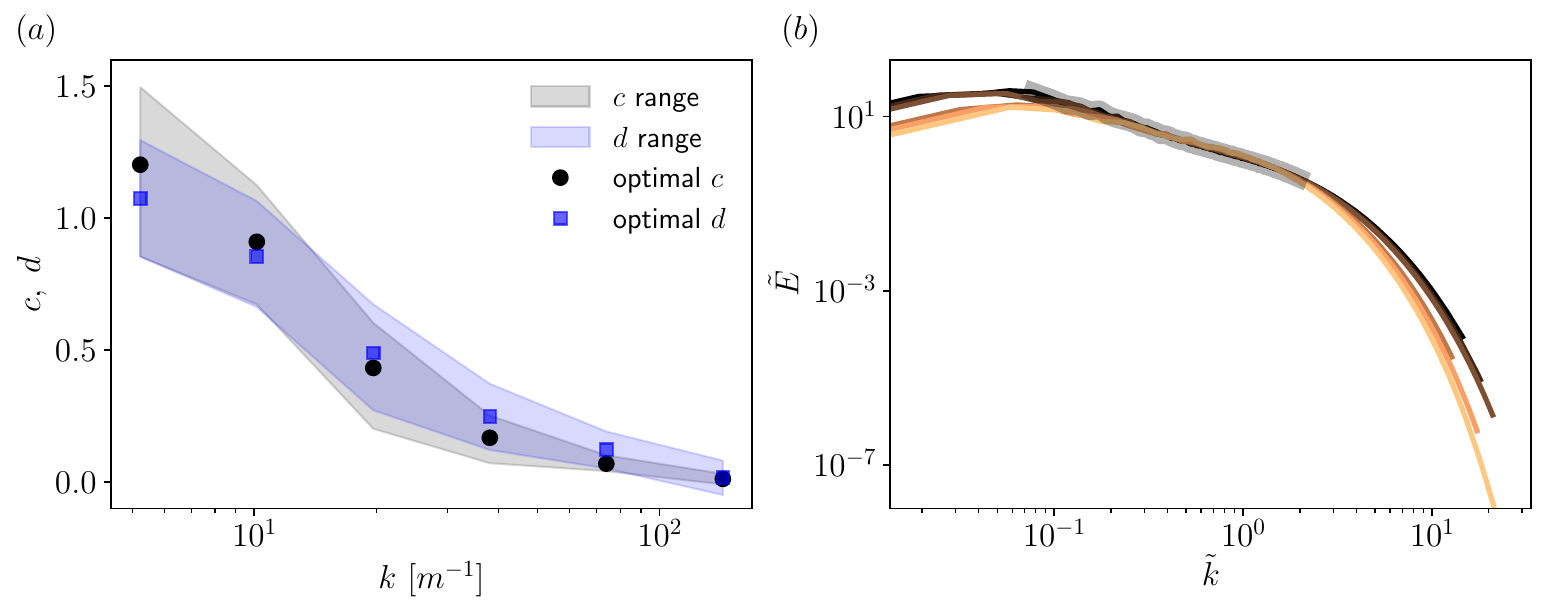}}
  \caption{$(a)$ Variation of the exponents $c$ (black circles) and $d$ (blue squares), as defined in \eqref{eq-dt-intermediate}, with the location of a fixed-length sliding window. The shaded areas highlight the admissible range of $c$ and $d$ values where $J<1.5J_m$ for each window. The window length is 1.42 decades. $(b)$ Rescaled energy spectra $\tilde{E}(\tilde{\kappa})$ using the expression identified for window 2, for which $J_m=0.0116$, obtained at $(c,d)=(0.433,0.489)$. The shaded region shows the collapsed range of the window.}
  \label{fig-dt_ab_y_2}
\end{figure}

\begin{figure}
  \centerline{\includegraphics[width=\textwidth]{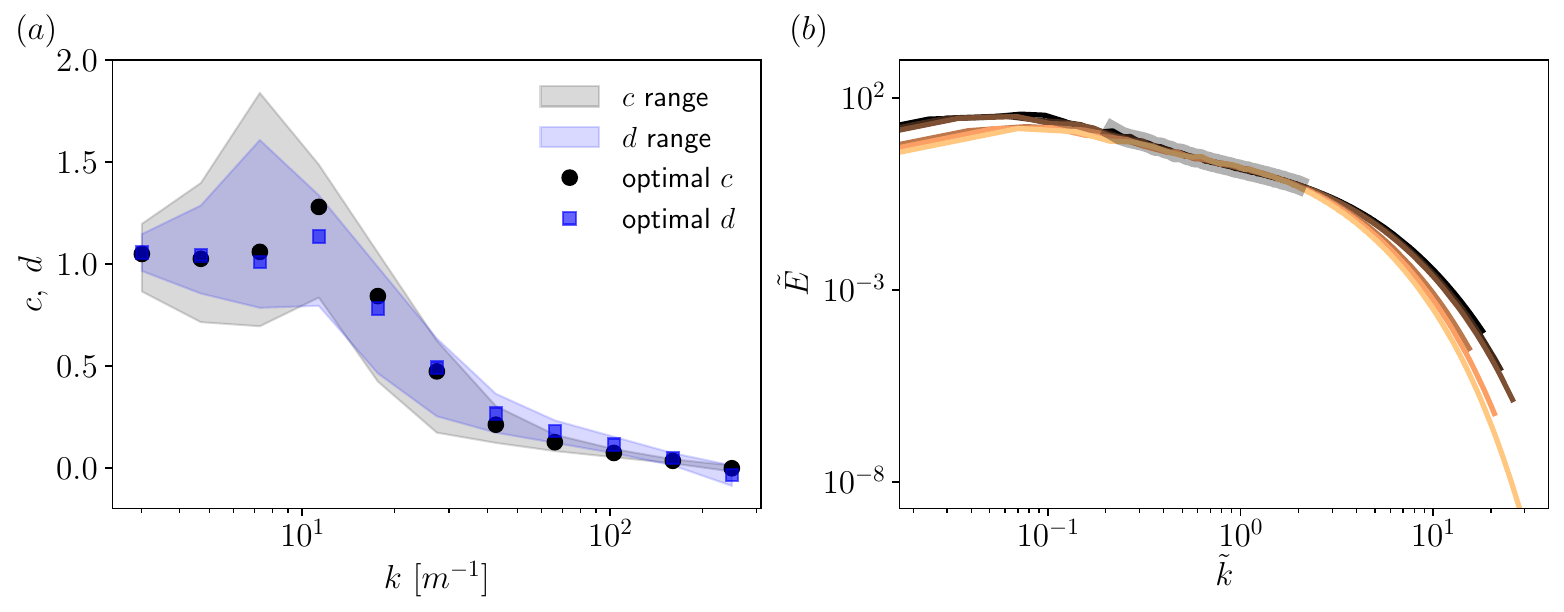}}
  \caption{$(a)$ Variation of the exponents $c$ (black circles) and $d$ (blue squares), as defined in \eqref{eq-dt-intermediate}, with the location of a fixed-length sliding window. The shaded areas highlight the admissible range of $c$ and $d$ values where $J<1.5J_m$ for each window. The window length is 1.07 decades. $(b)$ Rescaled energy spectra $\tilde{E}(\tilde{\kappa})$ using the expression identified for window 6, for which $J_m=0.0058$, obtained at $(c,d)=(0.473,0.491)$. The shaded region shows the collapsed range of the window.}
  \label{fig-dt_ab_y_3}
\end{figure}

For the case of late-stage homogeneous decaying turbulence, figures \ref{fig-dt_ab_y_2} and \ref{fig-dt_ab_y_3} show the variation of the exponents $c$ and $d$, defined in \eqref{eq-dt-intermediate}, as a function of the sliding window centre location, for different window lengths. The shaded areas show the admissible ranges of $(c, d)$ satisfying $J<1.5J_m$ for each window.

A monotonic variation of the exponents can be obtained by including cases with lower Reynolds numbers, as demonstrated in \S\ref{sec-ch-intermediate}, because the shorter overlap region imposes stronger constraints on the admissible values. Alternatively, increasing the window length can also produce a monotonic variation. 
This is shown in figures \ref{fig-dt_ab_y_2} and \ref{fig-dt_ab_y_3}, where the analysis is repeated with window lengths of 1.42 and 1.07 decades, respectively, corresponding to 2 and 1.5 times the length used in figure \ref{fig-dt_ab_y_4}.
With the increased window length, the admissible ranges of $c$ and $d$ satisfying $J<1.5J_m$ are reduced in the intermediate region, and the scatter in the optimal exponents is suppressed.
A monotonic increase in the exponents from 0 to 1 is observed in figure \ref{fig-dt_ab_y_2}$(a)$ as the window shifts from the inner to the outer region.
Windows 2 and 6 are chosen to show the collapse of the rescaled energy spectra in figures \ref{fig-dt_ab_y_2}$(b)$ and \ref{fig-dt_ab_y_3}$(b)$, respectively. As in figures \ref{fig-ch_ab_y_10}$(b)$ and \ref{fig-ch_ab_y_6}$(b)$, a good collapse is observed in the middle of the range, while it deteriorates toward the two ends.
These data-driven results again indicate the existence of an intermediate region in which the spectra can be collapsed using arbitrary values of $c$ and $d$, provided that $c \approx d$. This suggests that no characteristic length scale is distinguished within this region, consistent with the scale-free nature of the inertial subrange governed by the $-5/3$ law.

\section{Mathematical derivation of equation \eqref{eq-dt-intermediate-corr}}\label{app-dt-intermediate}

\begin{figure}
  \centerline{\includegraphics[width=0.5\textwidth]{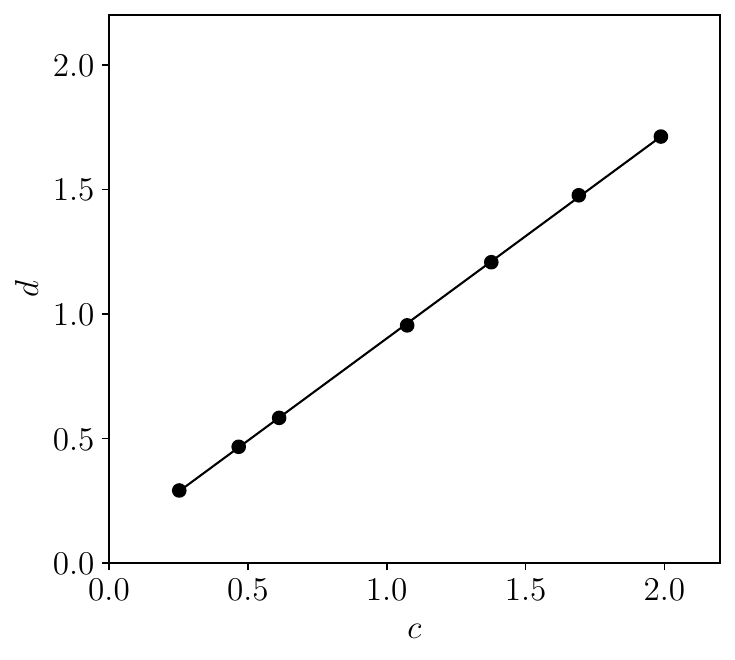}}
  \caption{Relationship between the exponents $c$ and $d$ in the intermediate range. The optimal values of $c$ and $d$ corresponding to windows 4 to 10 in figure \ref{fig-dt_ab_y_4} are employed in a linear fit, which gives $d = 0.8208c + 0.0809$ (black line).}
  \label{fig-dt_c_d_fit}
\end{figure}

We start from the following expression identified by the algorithm in the intermediate range,
\begin{gather}
  E = \epsilon^{2/3}(L^d\eta^{1-d})^{5/3}g(\kappa L^c\eta^{1-c}),\quad c\approx d \label{eq-dt-overlap}
\end{gather}
where $g$ is a nondimensional function, and $c, d$ can take arbitrary values. We consider two limiting cases: $c_0=0, d_0\approx0$ and $c_1=1, d_1\approx1$. Substituting the two limits of $c$ into \eqref{eq-dt-overlap} yields
\begin{gather}
  E = \epsilon^{2/3}(L^{d_0}\eta^{1-d_0})^{5/3}g_0(\kappa\eta), \label{eq-dt-c0d0}\\
  E = \epsilon^{2/3}(L^{d_1}\eta^{1-d_1})^{5/3}g_1(\kappa L). \label{eq-dt-c1d1}
\end{gather}
Equating them then leads to
\begin{gather}
  \left[\frac{(\kappa\eta)^{d_1}}{(\kappa\eta)^{d_0}}\right]^{5/3}g_0(\kappa\eta) = \left[\frac{(\kappa L)^{d_1}}{(\kappa L)^{d_0}}\right]^{5/3}g_1(\kappa L).
\end{gather}
Since the left- and right-hand sides depend only on $\kappa\eta$ and $\kappa L$, respectively, they must be equal to a constant. We thus obtain 
\begin{gather}
  g_0(K) = g_1(K) \propto \left(K^{d_0-d_1}\right)^{5/3}, \label{eq-dt-f0f1}
\end{gather}
where $K$ is the argument of the function. Substituting \eqref{eq-dt-f0f1} back into \eqref{eq-dt-c0d0} or \eqref{eq-dt-c1d1} then gives
\begin{gather}
  E \propto \epsilon^{2/3}\kappa^{-5/3}\left[\left(\kappa L\right)^{d_0}\left(\kappa\eta\right)^{1-d_1}\right]^{5/3}, \label{eq-dt-overlap-1}
\end{gather}
where $d_0$ and $d_1$ can be determined from the relation between $c$ and $d$, obtained from the linear fit of the computed values that yield a good collapse in the intermediate range. Here, we choose the data from the shortest window (figure \ref{fig-dt_ab_y_4}). The values of optimal $c$ and $d$ corresponding to windows 4 to 10 are then employed in a linear fit, as shown in figure \ref{fig-dt_c_d_fit}, which gives
\begin{gather}
  d = 0.8208c + 0.0809.
\end{gather}
Equation \eqref{eq-dt-overlap-1} then becomes
\begin{gather}
  E \propto \epsilon^{2/3}\kappa^{-5/3}\left[\left(\kappa L\right)^{0.0809}\left(\kappa\eta\right)^{0.0983}\right]^{5/3}, \label{eq-dt-overlap-2}
\end{gather}
which is equation \eqref{eq-dt-intermediate-corr}.

\bibliographystyle{jfm}
\bibliography{jfm}

\end{document}